\documentclass{article} 
\usepackage{authblk}
\usepackage{amsmath,amssymb}
\usepackage[dvips]{graphicx}
\usepackage{booktabs}
\usepackage{array}
\usepackage{comment}
\usepackage{authblk}
\usepackage{color}
\usepackage{psfrag}
\usepackage{soul}
\usepackage{enumerate}
\usepackage{caption}
\usepackage[normalem]{ulem}
\usepackage[super]{natbib}
\title{\bf One-sided Shewhart control charts for monitoring the ratio of two normal variables in Short Production Runs}
\author[1]{K.D. Tran}
\author[2]{Q.U.A Khaliq}
\author[4]{A. A. Nadi}
\author[3]{H Tran}
\author[4]{K.P. Tran\thanks{kim-phuc.tran@ensait.fr (corresponding author)}}
\affil[1]{Institute of Research and Development, Duy Tan University, Danang, 550000 Vietnam}
\affil[2]{Department of Statistics Allama Iqbal Open University Islamabad, Pakistan}
\affil[3]{Institute of Artificial Intelligence and Data Science, Dong A University, Danang, Vietnam}
\affil[4]{Ecole Nationale Sup\'erieure des Arts et Industries Textiles, GEMTEX Laboratory, BP 30329 59056 Roubaix Cedex 1, France}

\begin{document}
\maketitle
\begin{abstract}
Monitoring the ratio of two normal random variables plays an important role in several manufacturing environments. For short production runs, however, the control charts assumed infinite processes cannot function effectively to detect anomalies. In this paper, we tackle this problem by proposing two one-sided Shewhart-type charts to monitor the ratio of two normal random variables for a finite horizon production. The statistical performance of the proposed charts is investigated using the truncated average run length as a performance measure in short production runs. In order to help the quality practitioner to implement these control charts, we have provided ready-to-use tables of the control limit parameters. An illustrative example from the food industry is given for illustration.

\end{abstract}

\section{Introduction}
\label{sec:introduction}

Quality control (QC) enables the manufacturer to produce high-quality products according to the needs of the customer. Its tool kit helps to improve product quality by minimizing product cost, increase the efficiency of the process by reducing product waste. Control charts are famous tools for monitoring the assignable causes,  they also tell us when corrective action must be taken or timely notify us when corrective action must be taken to improve the process behavior (see \citet{Shewhart1931}). With growing competition in customer markets, manufacturers extremely depend on the quality of their products and services for their survival. The short-run manufacturing production process is become quite common for achieving the satisfaction of the customers like job-shops, which are categorized by a high amount of flexibility and manufacture diversity. Therefore, the life cycle of the products is decreasing rapidly.  Nowadays, the production lines in several manufacturings and engineering processes have limited.  In short runs production, some sources are fixed or the time span of the product is too short, maybe one hour or day. For example, any warehouse which may be on the lease. This warehouse is operating in the short run because it has a limited place. The owner cannot extend their business or shifted it to another place. When the agreement expires, he will have in position to enlarge business or shift to a large place. The same is the case, in manufacturing industries, like robotics manufacturing industry incorporates the limited production runs of automatic parts within the flexible production cells and semiconductor industry assembly of electronic boards and in beverages industries where the high volume of production and filling of soft drinks in every $24$ to $48$ hour required the $20$ to $30$ inspection between consecutive session.  \\
  
There is a myth among some manufacturing organizations, as they mostly feel that control charts are less applicable for the short-run process as the duration of their production cycle is too short. Recent studies signify control charts for the production process draw the attention of quality practitioners. Quality professionals are more concerned about the quality of the product. They are doing continuous improvement in product quality by reducing variability. The repute of any industry depends on the quality of goods or services that they offer to the customer. Quality is a major strategy that increases the productivity of any industry. \\

There are various SPC  (statistical process control) short run control charts  presented in literature to serve the purpose.  Short run control charts are more effective and useful for the small lot manufacturing runs with limited production data. \citet{ladany1973optimal} first introduced  optimized-p chart for short production run. Later on, numerous authors  introduced  effective designs  for such economic process,  see for instance  \citet{ladany1976selection}, \citet{del1993optimal}, \citet{del1996general}, \citet{tagaras1996dynamic}, \citet{nenes2006economically}, \citet{nenes2010evaluation}, \citet{celano2012economic}, \citet{castagliola2013variable}, \citet{castagliola2015one}, \citet{amdouni2015monitoring}, \citet{khatun2019one}, \citet{naseri2020statistical} and \citet{khoo2020median}. \\

This study is planned to design the Shewhart control chart for monitoring the ratio of two variables for a short production run. There are several production or manufacturing processes, where the quality characteristics of interest formed the ratio of two normal variables.  Production strategies, where numerous components need to be blended together to get a product composition can require monitoring the ratios of random variables when quality experts are frequently interested in the relative evaluation of the same property for two-components. In fact, guaranteeing the stable ratio between different components permits the product specifications to be encountered. Automotive, aerospace, electronics, pharmaceutical, materials production, food preparation, and packaging industries are typical applications of these manufacturing environments.  \\

\citet{celano2016design} designed the ratio type Shewhart control chart by using the data from the food processing company. They took the ratio of two seeds (pumpkin and flex) to meet the nutrition facts. \citet{tran2016monitoring} investigated the performance of the ratio chart using the run-rules scheme.   \citet{tran2016monitoring} designed EWMA type control charts for efficient monitoring of the ratio. \citet{tran2016performance} assessed $RZ$ chart efficiency and performance in presence of measurement error. \citet{tran2018steady} did an analysis to evaluate the steady-state ARL performance of the $RZ$ chart. \citet{tran2018monitoring} designed the CUSUM control chart which was more efficient than its $RZ$ setup. \citet{nguyen2019monitoring}   and  \citet{nguyen2020cusum} introduced EWMA and CUSUM design for ratio type control chart at several sampling intervals. \citet{du2020variable} applied the variable sampling interval idea to $RZ$ chart to evaluate its efficacy and performance. \citet{du2020effect} presented the one-sided $RZ$ chart which was more efficient in case of measurement error present in the process. \\

The articles cited above are planned to observe the ratio over a production horizon considered as infinite. However, there are numerous situations, where the production run no longer is infinite. As far as we know, no research has been done regarding the observing of the ratio (of two normal variables) infinite horizon framework. The aim of this study is to fill this gap by introducing two one-sided $RZ$  charts aiming to observe the decrease or increase in the finite runs.  In the remainder of the study, they are denoted by ``${Sh}^{+}_{RZ}$ Chart" for the upper sided chart and by ``${Sh}^{-}_{RZ}$ Chart" for the lower sided one. \\

The rest of the article organized as follows,  Section 2 denotes the main properties of ratio distribution.  Section 3 deals with the design structure of one-sided $RZ$ (${Sh}^{+}_{RZ}$ and ${Sh}^{-}_{RZ}$) Shewhart charts for a limited production run.    The truncated run length (\textit{TRL}), as the performance measure of the proposed short run charts, is briefly described in Section 4.   To assess the behavior of the new charts, a  numerical analysis is done in Section 5. Real-life illustration of the proposed technique presented in Section 6. Finally, conclusion and recommendation for future research are given in Section 7.\\
\section{Derivation of properties of the ratio distribution}
\label{sec:derivation}
Suppose that $X$ and $Y$ be the two normal random variables such that   $\mathbf{W}=(X, Y)^T\sim N(\pmb{\mu}_\mathbf{W},\pmb{\Sigma}_\mathbf{W}),$ i.e. $\mathbf{W}$ is a bivariate normal random vector with mean vector and variance-covariance matrix as follows:
\begin{align}\label{Eq1}
{\pmb{\mu}}_{\mathbf{W}}=\begin{pmatrix}
\mu_{X} \\ 
\mu_{Y}
\end{pmatrix} \quad \text{and}  \quad {\pmb{\Sigma}}_{\mathbf{W}}=\begin{pmatrix}
\sigma^2_{X} & \rho \sigma_{X}\sigma_{Y} \\ 
\rho \sigma_{X}\sigma_{Y} & {\sigma}^2_{Y} 
\end{pmatrix}  
\end{align} 
where $\mu_{X}$ and $\mu_{Y}$  are the means of two variables and $\rho$ is the correlation coefficient between them.  Coefficients of variation $\left(\gamma_{X},\gamma_{Y}\right)$ and standard-deviation ratio ($\omega$) of $X$   and  $Y$ are denoted by $\gamma_{X}=\frac{\sigma_{X}}{\mu_{Y}}$, $\gamma_{Y}=\frac{\sigma_{Y}}{\mu_{Y}}$ and $\omega=\frac{\sigma_{X}}{\sigma_{Y}}$, respectively. Let $Z$ be the ratio of $X$ to $ Y\left(Z={X}/{Y}\right)$. \citet{celano2016design}  derived an adequate approximation for the \textit{c.d.f} (cumulative distribution function) of $Z$  as a function of $\gamma_{X},\gamma_{Y},\omega,$ and $\rho$ as:
\begin{align}\label{Zcdf}
F_{Z}(z|\gamma_{X},\gamma_{Y},\omega,\rho) \simeq \Phi \left(\frac{A}{B}\right),
\end{align}
where
\begin{align*}
A=\frac{z}{\gamma_{Y}}-\frac{\omega}{\gamma_{X}}, \qquad \text{and} \qquad B=\sqrt{\omega^2-2\rho\omega z+{z}^2},
\end{align*}
and $\Phi$ is  the \textit{c.d.f} of standard normal distribution (Sdn). After some tedious derivations,  approximated \textit{p.d.f} (probability density function)  of the ratio $Z$  is 
\begin{align}\label{Zpdf}
f_{Z}(z|\gamma_{X},\gamma_{Y},\omega,\rho)\simeq \left(\frac{1}{B\gamma_{Y}}-\frac{\left(z-\rho\omega\right)A}{B^3}\right)\times \phi\left(\frac{A}{B}\right), 
\end{align}
where $\phi(\cdot)$ is the p.d.f. of Sdn. Solving the equation $F_{\mathrm{Z}}(z\left|\gamma_{X},\gamma_{Y},\omega,\rho\right)=p$  allows obtaining an approximate expression for the \textit{i.d.f} (inverse distribution function) $F^{-1}_{\mathrm{Z}}(p|\gamma_{X},\gamma_{Y},\omega,\rho)$. We have
\begin{align}\label{Zidf} 
F^{-1}_{Z}(p|\gamma_{X},\gamma_{Y},\omega,\rho)\simeq \begin{cases}
\frac{-C_2-\sqrt{C^2_2-4C_1C_3}}{2C_1}, \quad \text{if} \quad p \in (0,0.5], \\ 
\frac{-C_2+\sqrt{C^2_2-4C_1C_3}}{2C_1},\quad \text{if} \quad p \in [0.5,1), 
\end{cases} 
\end{align}
where $C_1$, $C_2$, and $C_3$ are functions of $\omega, \rho, \gamma_{X}, \gamma_{Y}, p$ and  they  are:
\begin{align*}
C_1&=\frac{1}{\gamma^2_{Y}}-\left(\Phi^{-1}(p)\right)^2, \\ 
C_2&=2\omega\left(\rho\left(\Phi^{-1}(p)\right)^2-\frac{1}{\gamma_{X}\gamma_{Y}}\right) \\ 
C_3&=\omega^2\left(\frac{1}{\gamma^2_{X}}-{\left({\Phi}^{-1}\left(\mathrm{p}\right)\right)}^2\right),
\end{align*}
and where $\Phi^{-1}(\cdot)$ is the \textit{i.d.f.} of Sdn.



\section{Design of two one sided Shewhart $RZ$ charts for short production Runs} 
\label{sec:one sided shewhart}
The production run is planned to produce small size lot having $N$ parts after a fixed rolling length ${H}$. Let $I$ be the number of planned inspections of rolling horizon $H$ and assume that no inspection takes place at the end of the run.    By these settings, the sampling frequency  (the time interval between two consecutive
inspections) will be $\mathcal{S}_h={{\frac{{H}}{\left(I+1\right)}}}$ hours.  The observed values of random variables $X$ and $Y$ are used to calculate the ratio for two one sided Shewhart charts for short run. In order to monitor the ratio   $Z$,   samples   of size $n$  will be taken at every sampling interval from the process   and  the quality characteristic $\mathbf{W}$ will be measured for each item. Let $[\mathbf{W}_{i,1}, \mathbf{W}_{i,2},\dots,\mathbf{W}_{i,n}]$ be the collected sample in which the couples $\mathbf{W}_{i,j}=(X_{i,j},Y_{i,j})^T$ for $j=1,2,...n$  follow the bivariate normal model $N(\pmb{\mu}_{\mathbf{W},i},\pmb{\Sigma}_{\mathbf{W},i})$  where:
\begin{align}\label{Eq7}
{\pmb{\mu}}_{\mathbf{W},i}=\begin{pmatrix}
\mu_{X,i} \\ 
\mu_{Y,i}
\end{pmatrix} \quad \text{and}  \quad {\pmb{\Sigma}}_{\mathbf{W},i}=\begin{pmatrix}
\sigma^2_{X,i} & \rho \sigma_{X,i}\sigma_{Y,i} \\ 
\rho \sigma_{X,i}\sigma_{Y,i} & {\sigma}^2_{Y,i} 
\end{pmatrix},  \quad i=1,2,3,...
\end{align}

To design the charts, let $\gamma_{X}$ and $\gamma_{Y}$ be the known and constant coefficients of variation and let  $z_0=\frac{\mu_{X,i}}{\mu_{Y,i}}$ and $\rho_{0}$  be the known in-control values of the ratio and the coefficient of correlation  that will ensure the stability of process. We further assume that linear relationships are held between the sample standard deviation and the sample mean for both variables $X$ and $Y$, i.e., $\sigma_{X,i}=\gamma_{X} \mu_{X,i}$ and $\sigma_{Y,i}=\gamma_{Y} \mu_{Y,i}$ for every $i \geq 1$. This  implies that    the  standard deviation  of each sample  can change proportionally to its mean  such that their ratio remains  constant. There are several quality characteristics in practice (such as weights, tensile strengths and linear dimensions) that  can have a dispersion proportional to the population mean. Finally, the sample units are considered free to vary from sample to sample and thus it is possible to have $\pmb{\mu}_{\mathbf{W},i} \neq \pmb{\mu}_{\mathbf{W},k}$  and $\pmb{\Sigma}_{\mathbf{W},i}\neq \pmb{\Sigma}_{\mathbf{W},k}$, for $i\neq k.$ The monitoring statistic of the proposed chart is the ratio of sample means that should be calculated at the inspections $i=1,2,,3,...$ as:
\begin{align}
\hat{Z}_{i}=\frac{\hat{\mu}_{X,i}}{\hat{\mu}_{Y,i}}=\frac{\bar{X}_{i}}{\bar{Y}_{i}}=\frac{\sum_{j=1}^{n}X_{i,j}}{\sum_{j=1}^{n}Y_{i,j}}
\end{align}
 
 To obtain the \textit{c.d.f} and \textit{i.d.f} of the statistic $\hat{Z}_{i}$, one needs to some distributional proportions of the sample means $\bar{X}_{i}$ and $\bar{Y}_{i}$. It is easy to see that $\bar{X}_{i}\sim N(\mu_{X,i},\frac{\sigma_{X,i}}{\sqrt{n}})$ and 
 $\bar{Y}_{i}\sim N(\mu_{Y,i},\frac{\sigma_{Y,i}}{\sqrt{n}})$ having the constant coefficients of variation $\gamma_{\bar{X}}=\frac{\gamma_{X}}{\sqrt{n}}$ and $\gamma_{\bar{Y}}=\frac{\gamma_{Y}}{\sqrt{n}}$. By the definition and the mentioned assumptions, the standard deviation ratio $\omega_{i}$ in each inspection can be calculated  as:
 \begin{align}
 \omega_{i}=\frac{\sigma_{X,i}}{\sigma_{Y,i}}=\frac{\mu_{X,i}}{\mu_{Y,i}} \frac{\gamma_{X}}{\gamma_{Y}}=z_{0} \times \frac{\gamma_{X}}{\gamma_{Y}}=\omega_{0},
 \end{align}
 where $\omega_{0}$ is   the in-control standard deviations ratio. We are now in a position to obtain the \textit{c.d.f}  and \textit{i.d.f} of $\hat{Z}_{i}$ based on the \textit{c.d.f} and \textit{i.d.f} of $Z$ in \eqref{Zcdf} and \eqref{Zidf}   as (\citet{du2020variable}):  
 \begin{align}
F_{\hat{Z}_{i}}(z|n,\gamma_{X},\gamma_{Y},z_{0},\rho_{0})&=F_{Z}\left(z|\frac{\gamma_{X}}{\sqrt{n}},\frac{\gamma_{Y}}{\sqrt{n}},\frac{z_{0}\gamma_{X}}{\gamma_{Y}},\rho_{0}\right),\\
F^{-1}_{\hat{Z}_{i}}(p|n,\gamma_{X},\gamma_{Y},z_{0},\rho_{0})&=F^{-1}_{Z}\left(p|\frac{\gamma_{X}}{\sqrt{n}},\frac{\gamma_{Y}}{\sqrt{n}},\frac{z_{0}\gamma_{X}}{\gamma_{Y}},\rho_{0}\right).
 \end{align}

 The control limits of two separate one sided ratio charts are as follows
\begin{enumerate}
\item  \textbf{One sided Shewhart-$R{Z}^{-}$ chart (${Sh}^{-}_{RZ}$ Chart)}
 The downward  ratio uses to monitor the decrease in ratio of two variables. The  lower and the upper control limits  $LCL^-$ and $UCL^-$   of said chart are:
 \begin{align}\label{CL-}
 LCL^-&=F^{-1}_{\hat{Z}_{i}}(\alpha_{0}|n,\gamma_{X},\gamma_{Y},z_{0},\rho_{0}) \\
 UCL^-&=+\infty.
 \end{align}
where $\alpha_{0}$ is the probability of type I error.
\item  \textbf{One sided Shewhart-$R{Z}^{+}$ chart ($Sh^{+}_{RZ}$ Chart)}
The upward   ratio uses to monitor the increase in ratio of two variables. The  lower and the upper control limits  $LCL^+$ and $UCL^+$   of said chart are:
\begin{align}\label{CL+}
 LCL^+&= 0\\
 UCL^+&=F^{-1}_{\hat{Z}_{i}}(1-\alpha_{0}|n,\gamma_{X},\gamma_{Y},z_{0},\rho_{0}).
\end{align}
\end{enumerate}

The ratio of two normal variables $\hat{Z}_{i}$ will be plotted against the limits. We will count all points which fall outside of the limits of both charts. The process is consider to be out-of-control (occ) if $\hat{Z}_{i}$ fulfills either of these conditions, $\hat{Z}_{i}<LCL^- \, \text{or} \, \hat{Z}_{i}>UCL^+$ and assignable causes will be  removed. As the production run is too small, the traditional run length will not be applicable to determine the run length properties of the control charts.   To serve the purpose truncated run-length \textit{TRL} and its average  (\textit{TARL}) will be used to determine the performance of the charts. The details and computations of  \textit{TRL} and \textit{TARL} are  provided in what follows.
\section{Properties of Truncated Run Length \textit{TRL}}
\label{sec:TRL}
In certain manufacturing processes, it could not be practical to gather a sufficient amount of data at the beginning of the manufacturing process for trial limits in phase-I. For example, in the aerospace industry, the production rate of huge components can be finite. It takes too much time to gather enough samples for the applying control charts. On the other hand, it is often needed that the quality control procedure starts as early as possible since the rate of each product is high. In such circumstances, traditional quality charts are not as effective. It may need more samples to create control limits until they are precise adequate to screen the whole process.  Traditional run-length does not give the reliable monitoring of assignable cause for such a process in which a finite (or limited) number \textit{I}  of samples is available in the production horizon having finite length $H$.  Therefore, \textit{TRL} as a modified version of traditional run length is used to determine the behavior of the process short production runs. It is define as ``the number of samples until a shift is identified or until completion of process either happen first".   \textit{TRL} is a discreet random variable with support $TRL \in \{1,2,...,I,I+1\}$. The event $TRL=l$ for $l \in \{1,2,...,I\}$ means that an ooc signal is declared by the chart before the termination of the production run and $TRL=I+1$ expresses that the run is completed without detecting any shift in $I$ inspections. The \textit{p.m.f} (probability mass function) $f_{TRL}(l)$ and the \textit{c.d.f} $F_{TRL}(l)$ of \textit{TRL} can be derived  as follows:

\begin{align}\label{TRLpmf} 
f_{TRL}(l)=\begin{cases}
p(1-p)^{l-1}, \quad \text{if} \quad l=1,2,...,I, \\ 
(1-p)^{I},\quad  \quad \, \, \, \text{if} \quad l=I+1, 
\end{cases} 
\end{align}
and 
\begin{align}\label{TRLcdf} 
F_{TRL}(l)=\begin{cases}
1-(1-p)^l, \quad \text{if} \quad l=1,2,...,I, \\ 
1, \qquad  \qquad \quad \, \, \, \, \text{if} \quad l=I+1, 
\end{cases} 
\end{align}
where $p$ is  the probability of an occ signal.  Let  $\textit{TARL}$ be the expected value of \textit{TRL}  that can be used in place of the traditional \textit{ARL} to evaluate the performance of the control charts in finite horizon production.  Moreover, let $\textit{TARL}_{0}$ and $\textit{TARL}_{1}$  be the in-control and out-of-control values of \textit{TARL}, respectively. Substituting $p$ by   $\alpha$ ($\alpha$ is the probability of type I error) or $1-\beta$ ($\beta$ is the probability of type II error)   in \eqref{TRLpmf}  and taking expectation over the \textit{TRL} support,  $\textit{TARL}_{0}$ and $\textit{TARL}_{1}$ will be derived as (\citet{nenes2010evaluation}):
\begin{align}
TARL_{0}&=\sum^{I}_{l=1}  l (1-\alpha)^{l-1}\alpha+(I+1)(1-\alpha)^I=\frac{1-(1-\alpha)^{I+1}}{\alpha},\\
TARL_{1}&=\sum^{I}_{l=1} l \beta^{l-1}(1-\beta)+(I+1)\beta^I=\frac{1-\beta^{I+1}}{1-\beta}.
\end{align}

 The error probabilities $\alpha$ and $\beta$ can be calculated for the proposed charts  by: 
\begin{align}\label{alpha} 
\alpha =\begin{cases}
F_{\hat{Z}_{i}}(LCL^-|n,\gamma_{X},\gamma_{Y},z_{0},\rho_{0}) , \quad \quad \, \, \, \, \text{for $\mathcal{S}h^-_{RZ}$ Chart}   \\ 
1-F_{\hat{Z}_{i}}(UCL^+|n,\gamma_{X},\gamma_{Y},z_{0},\rho_{0}),  \quad  \text{for $\mathcal{S}h^+_{RZ}$ Chart} 
\end{cases} 
\end{align}
and
\begin{align}\label{beta} 
\beta =\begin{cases}
1-F_{\hat{Z}_{i}}(LCL^-|n,\gamma_{X},\gamma_{Y},z_{1},\rho_{1}) , \quad \text{for $\mathcal{S}h^-_{RZ}$ Chart}   \\ 
F_{\hat{Z}_{i}}(UCL^+|n,\gamma_{X},\gamma_{Y},z_{1},\rho_{1}),  \quad \quad  \, \, \, \, \text{for $\mathcal{S}h^+_{RZ}$ Chart}
\end{cases} 
\end{align}
where $z_{1}=\tau z_{0} \, (\tau>0)$ and $\rho_{1}$ are the out-of-control values of the ratio and the coefficient
of correlation, respectively. While  the values of shift size   $\tau\in (0,1)$ correspond to decrease of nominal ratio $z_{0}$,   the values of $\tau>1$ correspond to  an increase  of nominal ratio $z_{0}$.

\section{Numerical Performance Analysis}
\label{sec:Analysis}
In this section, we evaluate the statistical performance of the two one-sided ${Sh}^{-}_{RZ}$ and ${Sh}^{+}_{RZ}$ charts for the short production run. The performance of the chart is evaluated by the startup of the short run when a particular shift $(\tau)$ happens. To serve the purpose, the  \textit{TARL} metric is used as the performance measure instead of using traditional  \textit{ARL}. Initially,  let   ${TARL}_0=I$   when the process expected to be in control. Statistical performance has been calculated by considering the following parameters values.
\begin{itemize}
\item  $z_0$=1 (in-control ratio of two variables)  

\item  $\gamma_{X} \in \{0.01,0.2\} \, \text{and} \, \gamma_{Y}\in \{0.01,0.2\}$

\item  $\rho_0\in \{0.0,\pm0.4,\pm0.8\}$

\item  $n=\{1,5,7,10,15\}$

\item  $I=\{10,30,50\}$

\item  $\tau=\left\{ \begin{array}{c}
0.9,0.95,0.98,0.99;\ \ \ \ \ \   for\  {Sh}^{-}_{RZ}\ chart\ \  \\ 
1.01,1.02,1.05, 1.1\ \ \ \ \ for\ {Sh}^{+}_{RZ}\ chart\  \end{array}
\right.$
\end{itemize}
Tables \ref{tab:Table1}-\ref{tab:Table3} illustrate the probability-type control limits   $LCL^-$ and $UCL^+$ values designing such that $TARL_0=I$ (\citet{nenes2006economically}) by varying $\gamma_{X}, \gamma_{Y}, \rho_0$ and $n$ at $\tau=\tau_0$ (when process is in control) and $z_0=1$. Tables \ref{tab:Table4}-\ref{tab:Table9} exhibit the $TARL_{1}$ values for two charts (${Sh}^{-}_{RZ}$ and ${Sh}^{+}_{RZ}$) for $I=10, 30$ and $50$, when $\rho_1=\rho_0$ using $n\in \left\{1, 5, 7, 15, 30\right\}, \gamma_{X}=\gamma_{Y}, \gamma_{X}\neq \gamma_{Y}$ and  $\tau=\left\{0.9, 0.95, 0.98, 0.99, 1.01, 1.02, 1.05, 1.1\right\}$. Tables \ref{tab:Table10}-\ref{tab:Table15} demonstrate the $TARL_{1}$ values  for two charts (${Sh}^{-}_{RZ}$ and ${Sh}^{+}_{RZ}$) for $I=10, 30$ and $50$, when $\rho_1\neq \rho_0$ using $n\in \{1,5,7,15,30\},  \tau=\{0.9, 0.95, 0.98, 0.99, 1.01, 1.02, 1.05, 1.1\},\gamma_{X}=\gamma_{Y}$, and $\gamma_{X}\neq \gamma_{Y}$. 
\begin{small}
\begin{table}[htp]
  \centering
  \caption{Values of $(LCL^{-}, UCL^{+})$ for $z_0=1,\ \gamma_{X}\in\left\{0.01,0.2\right\},  \gamma_{Y}\in\left\{0.01, 0.2\right\},\ \rho_0\in\left\{0.0, \pm 0.4, \pm 0.8\right\},\ n=\left\{1, 5, 7, 10, 15\right\}$ and $TARL_0=10.$}
	\resizebox{0.99\textwidth}{!}{
    \begin{tabular}{ccccccccccccccc}
		\toprule
  $\gamma_{X}$ & &$\gamma_{Y}$  & &$\rho_0$ &&$n=1$  &&$n=5$  &&$n=7$ &&$n=10$ &&$n=15$ \\
		\midrule
$0.01$ &&$0.01$ && $-0.8$&& $0.9615$ && $0.9826$ && $0.9853$ && $0.9877$ && $0.9899$ \\
       &&       &&        && $1.0401$ && $1.0177$ && $1.0150$ && $1.0125$ && $1.0102$ \\
       &&        && $-0.4$ && $0.9660$ && $0.9846$ && $0.9870$ && $0.9891$ && $0.9911$ \\
       &&        &&        && $1.0352$ && $1.0156$ && $1.0132$ && $1.0110$ && $1.0090$ \\
       &&        && $ 0.0$ && $0.9712$ && $0.9870$ && $0.9890$ && $0.9908$ && $0.9925$ \\
       &&        &&        && $1.0297$ && $1.0132$ && $1.0111$ && $1.0093$ && $1.0076$ \\
       &&        && $ 0.4$ && $0.9776$ && $0.9899$ && $0.9915$ && $0.9929$ && $0.9942$ \\
       &&        &&        && $1.0229$ && $1.0102$ && $1.0086$ && $1.0072$ && $1.0059$ \\
       &&        && $ 0.8$ && $0.9870$ && $0.9942$ && $0.9951$ && $0.9959$ && $0.9966$ \\
       &&        &&        && $1.0132$ && $1.0059$ && $1.0050$ && $1.0041$ && $1.0034$ \\ \midrule
$0.20$ && $0.20$ && $-0.8$ && $0.4326$ && $0.7008$ && $0.7413$ && $0.7789$ && $0.8158$ \\
       &&        &&        && $2.3116$ && $1.4269$ && $1.3490$ && $1.2838$ && $1.2258$ \\
       &&        && $-0.4$ && $0.4754$ && $0.7306$ && $0.7678$ && $0.8021$ && $0.8356$ \\
       &&        &&        && $2.1034$ && $1.3687$ && $1.3025$ && $1.2467$ && $1.1967$ \\
       &&        && $ 0.0$ && $0.5313$ && $0.7668$ && $0.7997$ && $0.8299$ && $0.8591$ \\
       &&        &&        && $1.8821$ && $1.3042$ && $1.2505$ && $1.2049$ && $1.1640$ \\
       &&        && $ 0.4$ && $0.6108$ && $0.8139$ && $0.8409$ && $0.8655$ && $0.8890$ \\
       &&        &&        && $1.6373$ && $1.2287$ && $1.1892$ && $1.1555$ && $1.1249$ \\
       &&        && $ 0.8$ && $0.7508$ && $0.8878$ && $0.9047$ && $0.9199$ && $0.9343$ \\
       &&        &&        && $1.3319$ && $1.1264$ && $1.1053$ && $1.0871$ && $1.0703$ \\ \midrule
$0.01$ && $0.20$ && $-0.8$ && $0.6954$ && $0.8375$ && $0.8592$ && $0.8796$ && $0.8995$ \\
       &&        &&        && $1.7346$ && $1.2363$ && $1.1929$ && $1.1567$ && $1.1245$ \\
       &&        && $-0.4$ && $0.7010$ && $0.8405$ && $0.8619$ && $0.8818$ && $0.9014$ \\
       &&        &&        && $1.7207$ && $1.2319$ && $1.1893$ && $1.1537$ && $1.1222$ \\
       &&        && $ 0.0$ && $0.7068$ && $0.8436$ && $0.8645$ && $0.8841$ && $0.9033$ \\
       &&        &&        && $1.7067$ && $1.2274$ && $1.1856$ && $1.1508$ && $1.1198$ \\
       &&        && $ 0.4$ && $0.7127$ && $0.8467$ && $0.8673$ && $0.8864$ && $0.9053$ \\
       &&        &&        && $1.6925$ && $1.2228$ && $1.1819$ && $1.1477$ && $1.1174$ \\
       &&        && $ 0.8$ && $0.7188$ && $0.8500$ && $0.8701$ && $0.8888$ && $0.9073$ \\
       &&        &&        && $1.6781$ && $1.2181$ && $1.1781$ && $1.1446$ && $1.1149$ \\ \midrule
$0.20$ && $0.01$ && $-0.8$ && $0.5765$ && $0.8089$ && $0.8383$ && $0.8645$ && $0.8893$ \\
       &&        &&        && $1.4381$ && $1.1941$ && $1.1638$ && $1.1369$ && $1.1117$ \\
       &&        && $-0.4$ && $0.5812$ && $0.8118$ && $0.8408$ && $0.8667$ && $0.8911$ \\
       &&        &&        && $1.4266$ && $1.1898$ && $1.1603$ && $1.1340$ && $1.1094$ \\
       &&        && $ 0.0$ && $0.5859$ && $0.8147$ && $0.8434$ && $0.8690$ && $0.8930$ \\
       &&        &&        && $1.4149$ && $1.1854$ && $1.1567$ && $1.1311$ && $1.1070$ \\
       &&        && $ 0.4$ && $0.5908$ && $0.8178$ && $0.8461$ && $0.8713$ && $0.8950$ \\
       &&        &&        && $1.4032$ && $1.1810$ && $1.1531$ && $1.1281$ && $1.1046$ \\
       &&        && $ 0.8$ && $0.5959$ && $0.8209$ && $0.8488$ && $0.8736$ && $0.8969$ \\
       &&        &&        && $1.3912$ && $1.1765$ && $1.1493$ && $1.1251$ && $1.1022$ \\
	\bottomrule
    \end{tabular}%
		}
  \label{tab:Table1}%
\end{table}%
\end{small}
\begin{small}
\begin{table}[htbp]
  \centering
  \caption{Values of $(LCL^{-}, UCL^{+})$ for $z_0=1,\ \gamma_{X}\in\left\{0.01,0.2\right\}, \gamma_{Y}\in\left\{0.01, 0.2\right\},\ \rho_0\in\left\{0.0, \pm 0.4, \pm 0.8\right\},\ n=\left\{1, 5, 7, 10, 15\right\}$ and $TARL_0=30.$ }
	\resizebox{0.99\textwidth}{!}{
    \begin{tabular}{ccccccccccccccc}
		\toprule
   $\gamma_{X}$ &&$\gamma_{Y}$  &&$\rho_0$ &&$n=1$  &&$n=5$  &&$n=7$ &&$n=10$ &&$n=15$ \\
		\midrule
$0.01$ && $0.01$ && $-0.8$ && $0.9474$ && $0.9761$ && $0.9798$ && $0.9831$ && $0.9861$ \\
       &&        &&        && $1.0555$ && $1.0245$ && $1.0206$ && $1.0172$ && $1.0141$ \\
       &&        && $-0.4$ && $0.9534$ && $0.9789$ && $0.9821$ && $0.9850$ && $0.9878$ \\
       &&        &&        && $1.0488$ && $1.0215$ && $1.0182$ && $1.0152$ && $1.0124$ \\
       &&        && $ 0.0$ && $0.9605$ && $0.9821$ && $0.9849$ && $0.9873$ && $0.9897$ \\
       &&        &&        && $1.0411$ && $1.0182$ && $1.0153$ && $1.0128$ && $1.0105$ \\
       &&        && $ 0.4$ && $0.9693$ && $0.9861$ && $0.9883$ && $0.9902$ && $0.9920$ \\
       &&        &&        && $1.0317$ && $1.0141$ && $1.0119$ && $1.0099$ && $1.0081$ \\
       &&        && $ 0.8$ && $0.9821$ && $0.9920$ && $0.9932$ && $0.9943$ && $0.9954$ \\
       &&        &&        && $1.0182$ && $1.0081$ && $1.0068$ && $1.0057$ && $1.0047$ \\ \midrule
$0.20$ && $0.20$ && $-0.8$ && $0.2908$ && $0.6097$ && $0.6601$ && $0.7077$ && $0.7549$ \\
       &&        &&        && $3.4390$ && $1.6402$ && $1.5149$ && $1.4131$ && $1.3248$ \\
       &&        && $-0.4$ && $0.3318$ && $0.6457$ && $0.6929$ && $0.7369$ && $0.7802$ \\
       &&        &&        && $3.0136$ && $1.5486$ && $1.4433$ && $1.3570$ && $1.2817$ \\
       &&        && $ 0.0$ && $0.3888$ && $0.6904$ && $0.7330$ && $0.7724$ && $0.8106$ \\
       &&        &&        && $2.5722$ && $1.4484$ && $1.3642$ && $1.2947$ && $1.2336$ \\
       &&        && $ 0.4$ && $0.4761$ && $0.7500$ && $0.7859$ && $0.8185$ && $0.8498$ \\
       &&        &&        && $2.1005$ && $1.3333$ && $1.2725$ && $1.2218$ && $1.1767$ \\
       &&        && $ 0.8$ && $0.6473$ && $0.8467$ && $0.8699$ && $0.8907$ && $0.9103$ \\
       &&        &&        && $1.5449$ && $1.1811$ && $1.1495$ && $1.1227$ && $1.0986$ \\ \midrule
$0.01$ && $0.20$ && $-0.8$ && $0.6223$ && $0.7887$ && $0.8156$ && $0.8412$ && $0.8666$ \\
       &&        &&        && $2.3771$ && $1.3557$ && $1.2855$ && $1.2286$ && $1.1794$ \\
       &&        && $-0.4$ && $0.6292$ && $0.7926$ && $0.8191$ && $0.8441$ && $0.8691$ \\
			 &&        &&        && $2.3510$ && $1.3490$ && $1.2801$ && $1.2243$ && $1.1761$ \\
       &&        && $ 0.0$ && $0.6364$ && $0.7966$ && $0.8226$ && $0.8471$ && $0.8716$ \\
       &&        &&        && $2.3246$ && $1.3422$ && $1.2747$ && $1.2200$ && $1.1726$ \\
       &&        && $ 0.4$ && $0.6437$ && $0.8007$ && $0.8261$ && $0.8502$ && $0.8742$ \\
       &&        &&        && $2.2981$ && $1.3353$ && $1.2692$ && $1.2155$ && $1.1692$ \\
       &&        && $ 0.8$ && $0.6513$ && $0.8050$ && $0.8298$ && $0.8534$ && $0.8768$ \\
       &&        &&        && $2.2712$ && $1.3283$ && $1.2635$ && $1.2110$ && $1.1656$ \\ \midrule
$0.20$ && $0.01$ && $-0.8$ && $0.4207$ && $0.7376$ && $0.7779$ && $0.8139$ && $0.8479$ \\
       &&        &&        && $1.6069$ && $1.2679$ && $1.2260$ && $1.1888$ && $1.1540$ \\
       &&        && $-0.4$ && $0.4254$ && $0.7413$ && $0.7812$ && $0.8168$ && $0.8503$ \\
       &&        &&        && $1.5893$ && $1.2616$ && $1.2209$ && $1.1847$ && $1.1507$ \\
       &&        && $ 0.0$ && $0.4302$ && $0.7450$ && $0.7845$ && $0.8197$ && $0.8528$ \\
       &&        &&        && $1.5715$ && $1.2553$ && $1.2157$ && $1.1805$ && $1.1473$ \\
       &&        && $ 0.4$ && $0.4351$ && $0.7489$ && $0.7879$ && $0.8227$ && $0.8553$ \\
       &&        &&        && $1.5535$ && $1.2488$ && $1.2105$ && $1.1762$ && $1.1439$ \\
       &&        && $ 0.8$ && $0.4403$ && $0.7528$ && $0.7914$ && $0.8257$ && $0.8579$ \\
       &&        &&        && $1.5354$ && $1.2423$ && $1.2051$ && $1.1718$ && $1.1405$ \\
	\bottomrule
    \end{tabular}%
  \label{tab:Table2}%
	}
\end{table}%
\end{small}
\begin{small}
\begin{table}[htbp]
  \centering
  \caption{Values of $(LCL^{-}, UCL^{+})$ for $z_0=1,\ \gamma_{X}\in\left\{0.01,0.2\right\}, \gamma_{Y}\in\left\{0.01, 0.2\right\},\ \rho_0\in\left\{0.0, \pm 0.4, \pm 0.8\right\},\ n=\left\{1, 5, 7, 10, 15\right\}$ and $TARL_0=50.$}
	\resizebox{0.99\textwidth}{!}{
    \begin{tabular}{ccccccccccccccc}
		\toprule
   $\gamma_{X}$ &&$\gamma_{Y}$  &&$\rho_0$ &&$n=1$  &&$n=5$  &&$n=7$ &&$n=10$ &&$n=15$ \\
		\midrule
$0.01$ && $0.01$ && $-0.8$ && $0.9418$ && $0.9736$ && $0.9776$ && $0.9812$ && $0.9846$ \\
       &&        &&        && $1.0618$ && $1.0272$ && $1.0229$ && $1.0191$ && $1.0156$ \\
       &&        && $-0.4$ && $0.9485$ && $0.9766$ && $0.9802$ && $0.9834$ && $0.9864$ \\
       &&        &&        && $1.0543$ && $1.0239$ && $1.0202$ && $1.0169$ && $1.0137$ \\
       &&        && $ 0.0$ && $0.9563$ && $0.9802$ && $0.9833$ && $0.9860$ && $0.9885$ \\
       &&        &&        && $1.0457$ && $1.0202$ && $1.0170$ && $1.0142$ && $1.0116$ \\
       &&        && $ 0.4$ && $0.9660$ && $0.9846$ && $0.9870$ && $0.9891$ && $0.9911$ \\
       &&        &&        && $1.0352$ && $1.0156$ && $1.0132$ && $1.0110$ && $1.0090$ \\
       &&        && $ 0.8$ && $0.9802$ && $0.9911$ && $0.9925$ && $0.9937$ && $0.9949$ \\
       &&        &&        && $1.0202$ && $1.0090$ && $1.0076$ && $1.0063$ && $1.0052$ \\ \midrule
$0.20$ && $0.20$ && $-0.8$ && $0.2411$ && $0.5760$ && $0.6298$ && $0.6809$ && $0.7317$ \\
       &&        &&        && $4.1479$ && $1.7361$ && $1.5877$ && $1.4687$ && $1.3666$ \\
       &&        && $-0.4$ && $0.2794$ && $0.6139$ && $0.6647$ && $0.7122$ && $0.7590$ \\
       &&        &&        && $3.5787$ && $1.6288$ && $1.5045$ && $1.4042$ && $1.3175$ \\
       &&        && $ 0.0$ && $0.3341$ && $0.6614$ && $0.7076$ && $0.7503$ && $0.7920$ \\
       &&        &&        && $2.9931$ && $1.5120$ && $1.4133$ && $1.3328$ && $1.2627$ \\
       &&        && $ 0.4$ && $0.4210$ && $0.7253$ && $0.7645$ && $0.8003$ && $0.8346$ \\
       &&        &&        && $2.3753$ && $1.3787$ && $1.3080$ && $1.2496$ && $1.1982$ \\
       &&        && $ 0.8$ && $0.6007$ && $0.8303$ && $0.8561$ && $0.8791$ && $0.9008$ \\
       &&        &&        && $1.6648$ && $1.2044$ && $1.1680$ && $1.1375$ && $1.1101$ \\ \midrule
$0.01$ && $0.20$ && $-0.8$ && $0.5971$ && $0.7708$ && $0.7995$ && $0.8268$ && $0.8541$ \\
       &&        &&        && $2.7832$ && $1.4095$ && $1.3262$ && $1.2596$ && $1.2027$ \\
       &&        && $-0.4$ && $0.6045$ && $0.7751$ && $0.8032$ && $0.8300$ && $0.8568$ \\
       &&        &&        && $2.7492$ && $1.4018$ && $1.3201$ && $1.2548$ && $1.1989$ \\
       &&        && $ 0.0$ && $0.6121$ && $0.7794$ && $0.8070$ && $0.8333$ && $0.8596$ \\
       &&        &&        && $2.7151$ && $1.3940$ && $1.3139$ && $1.2498$ && $1.1951$ \\
       &&        && $ 0.4$ && $0.6200$ && $0.7839$ && $0.8109$ && $0.8366$ && $0.8624$ \\
       &&        &&        && $2.6807$ && $1.3861$ && $1.3076$ && $1.2448$ && $1.1911$ \\
       &&        && $ 0.8$ && $0.6281$ && $0.7884$ && $0.8149$ && $0.8401$ && $0.8653$ \\
       &&        &&        && $2.6460$ && $1.3780$ && $1.3011$ && $1.2397$ && $1.1871$ \\ \midrule
$0.20$ && $0.01$ && $-0.8$ && $0.3593$ && $0.7095$ && $0.7540$ && $0.7939$ && $0.8315$ \\
       &&        &&        && $1.6746$ && $1.2973$ && $1.2508$ && $1.2095$ && $1.1708$ \\
       &&        && $-0.4$ && $0.3637$ && $0.7134$ && $0.7575$ && $0.7970$ && $0.8341$ \\
       &&        &&        && $1.6542$ && $1.2902$ && $1.2450$ && $1.2048$ && $1.1671$ \\
       &&        && $ 0.0$ && $0.3683$ && $0.7174$ && $0.7611$ && $0.8001$ && $0.8368$ \\
       &&        &&        && $1.6337$ && $1.2830$ && $1.2392$ && $1.2001$ && $1.1633$ \\
       &&        && $ 0.4$ && $0.3730$ && $0.7215$ && $0.7648$ && $0.8033$ && $0.8395$ \\
       &&        &&        && $1.6130$ && $1.2757$ && $1.2332$ && $1.1953$ && $1.1595$ \\
       &&        && $ 0.8$ && $0.3779$ && $0.7257$ && $0.7686$ && $0.8067$ && $0.8424$ \\
       &&        &&        && $1.5921$ && $1.2683$ && $1.2272$ && $1.1903$ && $1.1556$ \\
	\bottomrule
    \end{tabular}%
		}
  \label{tab:Table3}%
\end{table}%
\end{small}
The findings of proposed design in Tables \ref{tab:Table1}-\ref{tab:Table3} are  as follows:
\begin{enumerate}
\item  \textbf{Influence of sample size ($n$) to chart performance}
The width of control limits decreases as sample size ($n$) increases for  particular values of $\gamma_{X}, \gamma_{Y}, \rho_0$ and $I$. For example, when $\left(\gamma_{X},\ \ \gamma_{Y}\right)=\left(0.01, 0.01\right),\ n=1, I=10$, and $\rho_0$=$\rho_1=-0.8$ we have $(LCL^-, UCL^+)=(0.9615, 1.0401)$ while for, $\ n=15$, we have $(LCL^-, UCL^+)=(0.9899,1.0102)$ (cf. Table \ref{tab:Table1}). Similar pattern can be observed for other values.

\item  \textbf{Effect of ${\gamma}_{X}$ and ${\gamma}_{Y}$ values to chart performance }
Mostly, the control limits $LCL^- \, \text{and} \, UCL^+$ do not hold symmetry around 1. However, symmetry approximated attained by increasing sample size ($n$) for  strong  correlation and smaller values of  $(\gamma_{X},\gamma_{Y})$. In general, $UCL^+\neq \frac{1}{LCL^-}$; when $\gamma_{X}=\gamma_{Y}$ than $UCL^+=\frac{1}{LCL^-}$ hold. For example, when $\gamma_{X}=\gamma_{Y}=0.01$, $\rho_0$=$\rho_1=-0.8, n=1$, and $I=10$ the values of $LCL^- \, \text{and} \, UCL^+$are $0.9615$ and $1.0401$ that hold  symmetry  $\left(\frac{1}{0.9615}=1.0401\right)$ (cf. Table \ref{tab:Table1}). Similar pattern may be observe for $I=30 \& 50 (\gamma_{X}=\gamma_{Y})$ and which is accordance to \citet{celano2016design}.
\item \textbf{Impact of \textit{I} to chart performance }
 The number of inspections $I$ affects the width of control limits. The values of $LCL^- \, \text{and} \, UCL^+$, when $\gamma_{X}=\gamma_{Y}=0.01$, $\rho_0=\rho_1=-0.8$ are $(0.9615, 1.0401)$ for $I=10$ (cf. Table \ref{tab:Table1}),  $(0.9474, 1.055)$ for  $I=30$ (cf. Table \ref{tab:Table2}), and $(0.9418, 1.0618)$  for $I=50$  (cf. Table \ref{tab:Table3}). The value of $LCL^- \, \text{and} \, UCL^+$ using $\gamma_{X}=0.01, \gamma_{Y}=0.2, n=1,\rho_0$=$\rho_1=-0.8$ are $(0.6954, 1.7346)$ for $I=10, (0.6223, 2.3771)$ for $I=30$, and $(0.5971, 2.7832) $ for $I=50$. Similar pattern of $LCL^- \, \text{and} \, UCL^+$ can be observed for other values. Number of inspection widens the control limits.\\  
\end{enumerate}
\begin{flushleft}
\begin{table}[htbp]
\centering
\caption{$TARL_{1}$ values of the one-sided Shewhart $RZ$ chart ($Sh^{-}_{RZ}$ and $Sh^{+}_{RZ}$) for $z_0=1, \gamma_{X}\in \left\{0.01, 0.2\right\}, \gamma_{Y} \in \left\{0.01, 0.2\right\},\ \gamma_{X}=\gamma_{Y},\ \rho_0 \in \left\{-0.8, -0.4, 0.0,0.4,0.8\right\}, \rho_1=\rho_0, n=\left\{1, 5, 7, 10, 15\right\}, \tau=\left\{0.9, 0.95, 0.98, 0.99, 1, 1.01, 1.02, 1.05, 1.1 \right\}$ and $TARL_{0}=10$}
\resizebox{1.0\textwidth}{!}{
\begin{tabular}{l@{\hskip 0.2in}r@{\hskip 0.2in}r@{\hskip .2in}r@{\hskip .2in}r@{\hskip .2in}r@{\hskip .2in}r@{\hskip .2in}r@{\hskip .2in}r@{\hskip .2in}r@{\hskip .2in}r@{\hskip .2in}r@{\hskip .2in}r}\hline
& &\multicolumn{5}{c}{$(\gamma_{X}, \gamma_{Y})=(0.01, 0.01)$} & \phantom{abc}& \multicolumn{5}{c}{$(\gamma_X, \gamma_Y)=(0.2, 0.2)$} \\
\cmidrule{3-7} \cmidrule{9-13} 
$\tau$ && $n=1$ & $n=5$ & $n=7$ & $n=10$ & $n=15$ &&$n=1$ & $n=5$ & $n=7$ & $n=10$ & $n=15$\\ [6pt]\hline
& & \multicolumn{11}{c}{$\rho_0=\rho_1=-0.8$} \\[4pt] \hline
$0.90$ &  & $  1.0$ & $  1.0$ & $  1.0$ & $  1.0$ & $  1.0$ &  & $  9.3$ & $  7.8$ & $  7.2$ & $  6.4$ & $  5.4$ \\
$0.95$ &  & $  1.4$ & $  1.0$ & $  1.0$ & $  1.0$ & $  1.0$ &  & $  9.7$ & $  9.1$ & $  8.9$ & $  8.6$ & $  8.2$ \\
$0.98$ &  & $  5.4$ & $  1.6$ & $  1.3$ & $  1.1$ & $  1.0$ &  & $  9.9$ & $  9.7$ & $  9.6$ & $  9.6$ & $  9.4$ \\
$0.99$ &  & $  8.2$ & $  4.8$ & $  3.8$ & $  2.9$ & $  2.0$ &  & $  9.9$ & $  9.9$ & $  9.8$ & $  9.8$ & $  9.7$ \\
$1.01$ &  & $  8.2$ & $  4.8$ & $  3.9$ & $  2.9$ & $  2.1$ &  & $  9.9$ & $  9.9$ & $  9.8$ & $  9.8$ & $  9.7$ \\
\hline
\end{tabular}
\label{tab:Table4}
}
\end{table}
\end{flushleft}

\begin{flushleft}
\begin{table}[htbp]
\centering
\caption{$TARL_{1}$ values of the one-sided Shewhart $RZ$ chart ($Sh^{-}_{RZ}$ and $Sh^{+}_{RZ}$) for $z_0=1, \gamma_{X}\in \left\{0.01, 0.2\right\}, \gamma_{Y} \in \left\{0.01, 0.2\right\},\ \gamma_{X}\neq\gamma_{Y},\ \rho_0 \in \left\{-0.8, -0.4, 0.0, 0.4, 0.8\right\},\ \rho_1=\rho_0,\ n=\left\{1, 5, 7, 10, 15\right\},\ \tau=\left\{0.9, 0.95, 0.98, 0.99, 1, 1.01, 1.02, 1.05, 1.1 \right\}$ and $TARL_{0}=10.$}
\resizebox{1.0\textwidth}{!}{
\begin{tabular}{l@{\hskip 0.2in}r@{\hskip 0.2in}r@{\hskip .2in}r@{\hskip .2in}r@{\hskip .2in}r@{\hskip .2in}r@{\hskip .2in}r@{\hskip .2in}r@{\hskip .2in}r@{\hskip .2in}r@{\hskip .2in}r@{\hskip .2in}r}\hline
& &\multicolumn{5}{c}{$(\gamma_X, \gamma_Y)=(0.01, 0.2)$} & \phantom{abc}& \multicolumn{5}{c}{$(\gamma_X, \gamma_Y)=(0.2, 0.01)$}\\
\cmidrule{3-7} \cmidrule{9-13} 
$\tau$ && $n=1$ & $n=5$ & $n=7$ & $n=10$ & $n=15$ &&$n=1$ & $n=5$ & $n=7$ & $n=10$ & $n=15$\\ [6pt]\hline
& & \multicolumn{11}{c}{$\rho_0=\rho_1=-0.8$} \\[4pt] \hline
$0.90$ &  & $  7.5$ & $  4.4$ & $  3.5$ & $  2.7$ & $  2.0$ &  & $  9.0$ & $  5.8$ & $  4.7$ & $  3.5$ & $  2.4$ \\
$0.95$ &  & $  9.0$ & $  7.7$ & $  7.2$ & $  6.5$ & $  5.6$ &  & $  9.6$ & $  8.5$ & $  8.0$ & $  7.3$ & $  6.4$ \\
$0.98$ &  & $  9.7$ & $  9.3$ & $  9.1$ & $  9.0$ & $  8.7$ &  & $  9.9$ & $  9.5$ & $  9.4$ & $  9.2$ & $  9.0$ \\
$0.99$ &  & $  9.8$ & $  9.7$ & $  9.6$ & $  9.5$ & $  9.4$ &  & $  9.9$ & $  9.8$ & $  9.7$ & $  9.7$ & $  9.6$ \\
$1.01$ &  & $  9.9$ & $  9.8$ & $  9.7$ & $  9.7$ & $  9.6$ &  & $  9.8$ & $  9.7$ & $  9.6$ & $  9.5$ & $  9.4$ \\
\hline
\end{tabular}
}
\label{tab:Table5}
\end{table}
\end{flushleft}
\begin{flushleft}
\begin{table}[htbp]
\centering
\caption{$TARL_{1}$ values of the one-sided Shewhart $RZ$ chart ($Sh^{-}_{RZ}$ and $Sh^{+}_{RZ}$) for $z_0=1,\ \gamma_{X}\in \left\{0.01, 0.2\right\}, \gamma_{Y} \in \left\{0.01, 0.2\right\},\ \gamma_{X}=\gamma_{Y},\ \rho_0 \in \left\{-0.8, -0.4, 0.0, 0.4, 0.8\right\},\ \rho_1=\rho_0,\ n=\left\{1, 5, 7, 10, 15\right\},\ \tau=\left\{0.9, 0.95, 0.98, 0.99, 1, 1.01, 1.02, 1.05, 1.1 \right\}$ and $TARL_{0}=30.$}
\resizebox{1.0\textwidth}{!}{
\begin{tabular}{l@{\hskip 0.2in}r@{\hskip 0.2in}r@{\hskip .2in}r@{\hskip .2in}r@{\hskip .2in}r@{\hskip .2in}r@{\hskip .2in}r@{\hskip .2in}r@{\hskip .2in}r@{\hskip .2in}r@{\hskip .2in}r@{\hskip .2in}r}\hline
& &\multicolumn{5}{c}{$(\gamma_X, \gamma_Y)=(0.01, 0.01)$} & \phantom{abc}& \multicolumn{5}{c}{$(\gamma_X, \gamma_Y)=(0.2, 0.2)$} \\
\cmidrule{3-7} \cmidrule{9-13} 
$\tau$ && $n=1$ & $n=5$ & $n=7$ & $n=10$ & $n=15$ &&$n=1$ & $n=5$ & $n=7$ & $n=10$ & $n=15$\\ [6pt]\hline
& & \multicolumn{11}{c}{$\rho_0=\rho_1=-0.8$} \\[4pt]\hline
$0.90$ &  & $ 24.1$ & $ 13.4$ & $ 10.0$ & $  6.9$ & $  4.2$ &  & $ 29.0$ & $ 21.6$ & $ 17.6$ & $ 12.5$ & $  7.1$ \\
$0.95$ &  & $ 28.1$ & $ 25.1$ & $ 23.7$ & $ 21.7$ & $ 18.6$ &  & $ 29.6$ & $ 27.7$ & $ 26.7$ & $ 25.2$ & $ 22.6$ \\
$0.98$ &  & $ 29.4$ & $ 28.8$ & $ 28.6$ & $ 28.2$ & $ 27.7$ &  & $ 29.9$ & $ 29.4$ & $ 29.2$ & $ 28.9$ & $ 28.4$ \\
$0.99$ &  & $ 29.7$ & $ 29.5$ & $ 29.4$ & $ 29.3$ & $ 29.1$ &  & $ 29.9$ & $ 29.7$ & $ 29.6$ & $ 29.5$ & $ 29.4$ \\
$1.01$ &  & $ 29.9$ & $ 29.7$ & $ 29.6$ & $ 29.5$ & $ 29.4$ &  & $ 29.8$ & $ 29.5$ & $ 29.4$ & $ 29.3$ & $ 29.1$ \\
\hline
\end{tabular}
}
\label{tab:Table6}
\end{table}
\end{flushleft}
\begin{flushleft}
\begin{table}[htbp]
\centering
\caption{$TARL_{1}$ values of the one-sided Shewhart $RZ$ chart ($Sh^{-}_{RZ}$ and $Sh^{+}_{RZ}$) for $z_0=1, \gamma_{X}\in \left\{0.01, 0.2\right\}, \gamma_{Y} \in \left\{0.01, 0.2\right\}, \gamma_{X}\neq\gamma_{Y}, \rho_0 \in \left\{-0.8, -0.4, 0.0, 0.4, 0.8\right\}, \rho_1=\rho_0, n=\left\{1, 5, 7, 10, 15\right\}, \tau=\left\{0.9, 0.95, 0.98, 0.99, 1, 1.01, 1.02, 1.05, 1.1 \right\}$ and $TARL_{0}=30.$}
\resizebox{1.0\textwidth}{!}{
\begin{tabular}{l@{\hskip 0.2in}r@{\hskip 0.2in}r@{\hskip .2in}r@{\hskip .2in}r@{\hskip .2in}r@{\hskip .2in}r@{\hskip .2in}r@{\hskip .2in}r@{\hskip .2in}r@{\hskip .2in}r@{\hskip .2in}r@{\hskip .2in}r}\hline
& &\multicolumn{5}{c}{$(\gamma_X, \gamma_Y)=(0.01, 0.2)$} & \phantom{abc}& \multicolumn{5}{c}{$(\gamma_X, \gamma_Y)=(0.2, 0.01)$}\\
\cmidrule{3-7} \cmidrule{9-13} 
$\tau$ && $n=1$ & $n=5$ & $n=7$ & $n=10$ & $n=15$ &&$n=1$ & $n=5$ & $n=7$ & $n=10$ & $n=15$\\ [6pt]\hline
& & \multicolumn{11}{c}{$\rho_0=\rho_1=-0.8$} \\[4pt]\hline
$0.90$ &  & $ 24.1$ & $ 13.4$ & $ 10.0$ & $  6.9$ & $  4.2$ &  & $ 29.0$ & $ 21.6$ & $ 17.6$ & $ 12.5$ & $  7.1$ \\
$0.95$ &  & $ 28.1$ & $ 25.1$ & $ 23.7$ & $ 21.7$ & $ 18.6$ &  & $ 29.6$ & $ 27.7$ & $ 26.7$ & $ 25.2$ & $ 22.6$ \\
$0.98$ &  & $ 29.4$ & $ 28.8$ & $ 28.6$ & $ 28.2$ & $ 27.7$ &  & $ 29.9$ & $ 29.4$ & $ 29.2$ & $ 28.9$ & $ 28.4$ \\
$0.99$ &  & $ 29.7$ & $ 29.5$ & $ 29.4$ & $ 29.3$ & $ 29.1$ &  & $ 29.9$ & $ 29.7$ & $ 29.6$ & $ 29.5$ & $ 29.4$ \\
$1.01$ &  & $ 29.9$ & $ 29.7$ & $ 29.6$ & $ 29.5$ & $ 29.4$ &  & $ 29.8$ & $ 29.5$ & $ 29.4$ & $ 29.3$ & $ 29.1$ \\
\hline
\end{tabular}
}
\label{tab:Table7}
\end{table}
\end{flushleft}

\begin{flushleft}
\begin{table}[htbp]
\centering
\caption{$TARL_{1}$ values of the one-sided Shewhart $RZ$ chart ($Sh^{-}_{RZ}$ and $Sh^{+}_{RZ}$) for $z_0=1, \gamma_{X}\in \left\{0.01, 0.2\right\}, \gamma_{Y} \in \left\{0.01, 0.2\right\}, \gamma_{X}=\gamma_{Y}, \rho_0 \in \left\{-0.8, -0.4, 0.0, 0.4, 0.8\right\}, \rho_1=\rho_0, n=\left\{1, 5, 7, 10, 15\right\}, \tau=\left\{0.9, 0.95, 0.98, 0.99, 1, 1.01, 1.02, 1.05, 1.1 \right\}$ and $TARL_{0}=50.$}
\resizebox{1.0\textwidth}{!}{
\begin{tabular}{l@{\hskip 0.2in}r@{\hskip 0.2in}r@{\hskip .2in}r@{\hskip .2in}r@{\hskip .2in}r@{\hskip .2in}r@{\hskip .2in}r@{\hskip .2in}r@{\hskip .2in}r@{\hskip .2in}r@{\hskip .2in}r@{\hskip .2in}r}\hline
& &\multicolumn{5}{c}{$(\gamma_X, \gamma_Y)=(0.01, 0.01)$} & \phantom{abc}& \multicolumn{5}{c}{$(\gamma_X, \gamma_Y)=(0.2, 0.2)$} \\
\cmidrule{3-7} \cmidrule{9-13} 
$\tau$ && $n=1$ & $n=5$ & $n=7$ & $n=10$ & $n=15$ &&$n=1$ & $n=5$ & $n=7$ & $n=10$ & $n=15$\\ [6pt]\hline
& & \multicolumn{11}{c}{$\rho_0=\rho_1=-0.8$} \\[4pt]\hline
$0.90$ &  & $ 41.6$ & $ 23.2$ & $ 17.0$ & $ 11.0$ & $  6.3$ &  & $ 49.1$ & $ 39.3$ & $ 32.9$ & $ 23.5$ & $ 12.7$ \\
$0.95$ &  & $ 47.6$ & $ 43.4$ & $ 41.3$ & $ 38.2$ & $ 33.0$ &  & $ 49.6$ & $ 47.4$ & $ 46.1$ & $ 44.1$ & $ 40.3$ \\
$0.98$ &  & $ 49.3$ & $ 48.6$ & $ 48.3$ & $ 47.8$ & $ 47.1$ &  & $ 49.9$ & $ 49.3$ & $ 49.1$ & $ 48.7$ & $ 48.2$ \\
$0.99$ &  & $ 49.7$ & $ 49.4$ & $ 49.3$ & $ 49.2$ & $ 49.0$ &  & $ 49.9$ & $ 49.7$ & $ 49.6$ & $ 49.5$ & $ 49.3$ \\
$1.01$ &  & $ 49.9$ & $ 49.7$ & $ 49.6$ & $ 49.5$ & $ 49.3$ &  & $ 49.7$ & $ 49.4$ & $ 49.3$ & $ 49.2$ & $ 49.0$ \\
\hline
\end{tabular}
}
\label{tab:Table8}
\end{table}
\end{flushleft}

\begin{flushleft}
\begin{table}[htbp]
\centering
\caption{$TARL_{1}$ values of the one-sided Shewhart $RZ$ chart ($Sh^{-}_{RZ}$ and $Sh^{+}_{RZ}$) for $z_0=1, \gamma_{X}\in \left\{0.01, 0.2\right\}, \gamma_{Y} \in \left\{0.01, 0.2\right\}, \gamma_{X}\neq\gamma_{Y}, \rho_0 \in \left\{-0.8, -0.4, 0.0, 0.4, 0.8\right\}, \rho_1=\rho_0, n=\left\{1, 5, 7, 10, 15\right\}, \tau=\left\{0.9, 0.95, 0.98, 0.99, 1, 1.01, 1.02, 1.05, 1.1 \right\}$ and $TARL_{0}=50.$}
\resizebox{1.0\textwidth}{!}{
\begin{tabular}{l@{\hskip 0.2in}r@{\hskip 0.2in}r@{\hskip .2in}r@{\hskip .2in}r@{\hskip .2in}r@{\hskip .2in}r@{\hskip .2in}r@{\hskip .2in}r@{\hskip .2in}r@{\hskip .2in}r@{\hskip .2in}r@{\hskip .2in}r}\hline
& &\multicolumn{5}{c}{$(\gamma_X, \gamma_Y)=(0.01, 0.2)$} & \phantom{abc}& \multicolumn{5}{c}{$(\gamma_X, \gamma_Y)=(0.2, 0.01)$} \\
\cmidrule{3-7} \cmidrule{9-13} 
$\tau$ && $n=1$ & $n=5$ & $n=7$ & $n=10$ & $n=15$ &&$n=1$ & $n=5$ & $n=7$ & $n=10$ & $n=15$\\ [6pt]\hline
& & \multicolumn{11}{c}{$\rho_0=\rho_1=-0.8$} \\[4pt]\hline
$0.90$ &  & $ 41.6$ & $ 23.2$ & $ 17.0$ & $ 11.0$ & $  6.3$ &  & $ 49.1$ & $ 39.3$ & $ 32.9$ & $ 23.5$ & $ 12.7$ \\
$0.95$ &  & $ 47.6$ & $ 43.4$ & $ 41.3$ & $ 38.2$ & $ 33.0$ &  & $ 49.6$ & $ 47.4$ & $ 46.1$ & $ 44.1$ & $ 40.3$ \\
$0.98$ &  & $ 49.3$ & $ 48.6$ & $ 48.3$ & $ 47.8$ & $ 47.1$ &  & $ 49.9$ & $ 49.3$ & $ 49.1$ & $ 48.7$ & $ 48.2$ \\
$0.99$ &  & $ 49.7$ & $ 49.4$ & $ 49.3$ & $ 49.2$ & $ 49.0$ &  & $ 49.9$ & $ 49.7$ & $ 49.6$ & $ 49.5$ & $ 49.3$ \\
$1.01$ &  & $ 49.9$ & $ 49.7$ & $ 49.6$ & $ 49.5$ & $ 49.3$ &  & $ 49.7$ & $ 49.4$ & $ 49.3$ & $ 49.2$ & $ 49.0$ \\
\hline
\end{tabular}
}
\label{tab:Table9}
\end{table}
\end{flushleft}
The following results  can be drawn from Tables \ref{tab:Table4}-\ref{tab:Table9}.
\begin{enumerate}
\item  When shift size $\tau $ decrease or increase at fixed values of $I, n$ and $\rho_0$, the $TARL$ will decrease or increase accordingly. It can be observe that $TARL_1 < TARL_0$  in all cases. For instance, at fixed values of $I=10, \gamma_{X}=\gamma_{Y}=0.01, n=1, \rho_1=\rho_0=-0.8, TARL_0=10$ and for a specific $\tau$,  $TARL_{1}$ values  are $1, 1.4, 5.4, 8.2 \, \text{and} \, 8.2, 5.5, 1.4, 1$. Similar pattern of $TARL_{1}$ may be noted for other values of $I=30, 50,\gamma_{X}=\gamma_{Y}, \rho_1=\rho_0$. Since, the ratio distribution is skewed but one sided Shewhart charts attain symmetry.

\item  As sample size increases, the value of $TARL{1}$ decreases. Sample size influences the chart performance for specific  values of $\tau, I, \gamma_{X}=\gamma_{Y}$, $\rho_1=\rho_0$. For example,  $TARL_{1}$ values  are $8.2, 4.8, 3.9, 2.9, 2.0$ for  ${\mathcal{S}h}^-_{RZ} chart$ with $\tau=0.99$ and are $8.2, 4.8, 3.8, 2.9, 2.1$ for ${\mathcal{S}h}^+_{RZ} chart$ with $\tau=1.01$ at $n\in \left\{1, 5, 7, 15, 30\right\}$, $I=10$, and $\rho_1=\rho_0=-0.8$ (cf. Table-\ref{tab:Table4}). ${\mathcal{S}h}^-_{RZ}\ chart$ is more sensitive to shifts than ${\mathcal{S}h}^+_{RZ}\ chart$ in ratio $z$. Similar pattern may be observed for $I=30 \, \text{and} \, 50$.

\item  The charts performance is strongly influenced by values of $\gamma_{X}, \gamma_{Y}$ and $I.$ Proposed design is more efficient, when $\gamma_{X}=\gamma_{Y}$ at ${TARL}_0=10$. Both charts equally performed, when ${TARL}_0=30,50$, $\gamma_{X}=\gamma_{Y}$ and $\gamma_{X}\neq \gamma_{Y}$. For example, when ${TARL}_0=10, n=1, \rho_0=\rho_1=-0.8$, $\tau^+=1.01 \,\text{and} \,\tau^-=0.99$, we have ${TARL}_1$ is  $8.2$  for $\gamma_{X}=\gamma_{Y}=0.01$. However for ${\gamma}_{X}=.01\ \& \ \gamma_{Y}=0.2$, we have $9.8 \& 9.9$ (cf. Table (\ref{tab:Table4})). The performance of one sided Shewhart chart is identical ${TARL}_0=30\ \&\ 50$ for all values of $\gamma_{X}$\ and $\ \gamma_{Y}$ (cf. Table (\ref{tab:Table5})) . We have ${TARL}_0=30,\ n=1,\ \rho_0=\rho_1=-0.8,\tau=1.01,\ 0.99,$ we have ${TARL}_1=29.7$ and $29.9$ (for $\gamma_{X}=\gamma_{Y}=.01\ \& \gamma_{X}=.01\ ,\ \gamma_{Y}=0.2)$ (cf. Table (\ref{tab:Table4}-\ref{tab:Table9})). \\ \\ \\ \\ \\
\end{enumerate}
\begin{flushleft}
\begin{table}[htbp]
\centering
\caption{$TARL_{1}$ values of the one-sided Shewhart $RZ$ chart ($Sh^{-}_{RZ}$ and $Sh^{+}_{RZ}$) for $z_0=1, \gamma_{X}\in\left\{0.01, 0.2\right\}, \gamma_{Y}\in\left\{0.01, 0.2\right\}, \gamma_{X}=\gamma_{Y}, \rho_0\in\left\{-0.8, -0.4, 0.0, 0.4, 0.8\right\}, \rho_1=\rho_0, n=\left\{1, 5, 7, 10, 15\right\}, \tau=\left\{0.9, 0.95, 0.98, 0.99, 1, 1.01, 1.02, 1.05, 1.1 \right\}$ and $TARL_{0}=10.$}
\resizebox{1.0\textwidth}{!}{
\begin{tabular}{l@{\hskip 0.2in}r@{\hskip 0.2in}r@{\hskip .2in}r@{\hskip .2in}r@{\hskip .2in}r@{\hskip .2in}r@{\hskip .2in}r@{\hskip .2in}r@{\hskip .2in}r@{\hskip .2in}r@{\hskip .2in}r@{\hskip .2in}r}\hline
& &\multicolumn{5}{c}{$(\gamma_X, \gamma_Y)=(0.01, 0.01)$} & \phantom{abc}& \multicolumn{5}{c}{$(\gamma_X, \gamma_Y)=(0.2, 0.2)$} \\
\cmidrule{3-7} \cmidrule{9-13} 
$\tau$ && $n=1$ & $n=5$ & $n=7$ & $n=10$ & $n=15$ &&$n=1$ & $n=5$ & $n=7$ & $n=10$ & $n=15$\\ [6pt]\hline
& & \multicolumn{11}{c}{$\rho_0=-0.4,\rho_1=-0.2$} \\[4pt]\hline
$0.90$ &  & $  1.0$ & $  1.0$ & $  1.0$ & $  1.0$ & $  1.0$ &  & $  9.6$ & $  7.9$ & $  7.2$ & $  6.3$ & $  5.0$ \\
$0.95$ &  & $  1.2$ & $  1.0$ & $  1.0$ & $  1.0$ & $  1.0$ &  & $ 10.0$ & $  9.5$ & $  9.2$ & $  8.9$ & $  8.4$ \\
$0.98$ &  & $  5.0$ & $  1.3$ & $  1.1$ & $  1.0$ & $  1.0$ &  & $ 10.2$ & $ 10.0$ & $ 10.0$ & $  9.9$ & $  9.8$ \\
$0.99$ &  & $  8.4$ & $  4.3$ & $  3.2$ & $  2.3$ & $  1.6$ &  & $ 10.2$ & $ 10.2$ & $ 10.2$ & $ 10.1$ & $ 10.1$ \\
$1.00$ &  & $ 10.3$ & $ 10.3$ & $ 10.3$ & $ 10.3$ & $ 10.3$ &  & $ 10.3$ & $ 10.3$ & $ 10.3$ & $ 10.3$ & $ 10.3$ \\
\hline
\end{tabular}
}
\label{tab:Table10}
\end{table}
\end{flushleft}
\begin{flushleft}
\begin{table}[htbp]
\centering
\caption{$TARL_{1}$ values of the one-sided Shewhart $RZ$ chart ($Sh^{-}_{RZ}$ and $Sh^{+}_{RZ}$) for $z_0=1, \gamma_{X}\in \left\{0.01, 0.2\right\}, \gamma_{Y} \in \left\{0.01, 0.2\right\}, \gamma_{X}\neq\gamma_{Y}, \rho_0 \in \left\{-0.8, -0.4, 0.0, 0.4, 0.8\right\}, \rho_1\neq\rho_0, n=\left\{1, 5, 7, 10, 15\right\}, \tau=\left\{0.9, 0.95, 0.98, 0.99, 1, 1.01, 1.02, 1.05, 1.1 \right\}$ and $TARL_{0}=10.$}
\resizebox{1.0\textwidth}{!}{
\begin{tabular}{l@{\hskip 0.2in}r@{\hskip 0.2in}r@{\hskip .2in}r@{\hskip .2in}r@{\hskip .2in}r@{\hskip .2in}r@{\hskip .2in}r@{\hskip .2in}r@{\hskip .2in}r@{\hskip .2in}r@{\hskip .2in}r@{\hskip .2in}r}\hline
& &\multicolumn{5}{c}{$(\gamma_X, \gamma_Y)=(0.01, 0.2)$} & \phantom{abc}& \multicolumn{5}{c}{$(\gamma_X, \gamma_Y)=(0.2, 0.01)$} \\
\cmidrule{3-7} \cmidrule{9-13} 
$\tau$ && $n=1$ & $n=5$ & $n=7$ & $n=10$ & $n=15$ &&$n=1$ & $n=5$ & $n=7$ & $n=10$ & $n=15$\\ [6pt]\hline

& & \multicolumn{11}{c}{$\rho_0=-0.4,\rho_1=-0.2$} \\[4pt]\hline
$0.90$ &  & $  7.5$ & $  4.3$ & $  3.5$ & $  2.7$ & $  2.0$ &  & $  9.1$ & $  5.8$ & $  4.7$ & $  3.5$ & $  2.3$ \\
$0.95$ &  & $  9.0$ & $  7.7$ & $  7.2$ & $  6.5$ & $  5.5$ &  & $  9.6$ & $  8.5$ & $  8.0$ & $  7.3$ & $  6.4$ \\
$0.98$ &  & $  9.7$ & $  9.3$ & $  9.2$ & $  9.0$ & $  8.7$ &  & $  9.9$ & $  9.6$ & $  9.4$ & $  9.3$ & $  9.0$ \\
$0.99$ &  & $  9.9$ & $  9.7$ & $  9.7$ & $  9.6$ & $  9.5$ &  & $ 10.0$ & $  9.8$ & $  9.8$ & $  9.7$ & $  9.6$ \\
$1.00$ &  & $ 10.0$ & $ 10.0$ & $ 10.0$ & $ 10.0$ & $ 10.0$ &  & $ 10.1$ & $ 10.1$ & $ 10.1$ & $ 10.1$ & $ 10.1$ \\
\hline
\end{tabular}
}
\label{tab:Table11}
\end{table}
\end{flushleft}
\begin{flushleft}
\begin{table}[htbp]
\centering
\caption{$TARL_{1}$ values of the one-sided Shewhart $RZ$ chart ($Sh^{-}_{RZ}$ and $Sh^{+}_{RZ}$) for $z_0=1, \gamma_{X}\in \left\{0.01, 0.2\right\}, \gamma_{Y} \in \left\{0.01, 0.2\right\}, \gamma_{X}=\gamma_{Y}, \rho_0 \in \left\{-0.8, -0.4, 0.0, 0.4, 0.8\right\}, \rho_1\neq\rho_0, n=\left\{1, 5, 7, 10, 15\right\}, \tau=\left\{0.9, 0.95, 0.98, 0.99, 1, 1.01, 1.02, 1.05, 1.1 \right\}$ and $TARL_{0}=30.$}
\resizebox{1.0\textwidth}{!}{
\begin{tabular}{l@{\hskip 0.2in}r@{\hskip 0.2in}r@{\hskip .2in}r@{\hskip .2in}r@{\hskip .2in}r@{\hskip .2in}r@{\hskip .2in}r@{\hskip .2in}r@{\hskip .2in}r@{\hskip .2in}r@{\hskip .2in}r@{\hskip .2in}r}\hline
& &\multicolumn{5}{c}{$(\gamma_X, \gamma_Y)=(0.01, 0.01)$} & \phantom{abc}& \multicolumn{5}{c}{$(\gamma_X, \gamma_Y)=(0.2, 0.2)$} \\
\cmidrule{3-7} \cmidrule{9-13} 
$\tau$ && $n=1$ & $n=5$ & $n=7$ & $n=10$ & $n=15$ &&$n=1$ & $n=5$ & $n=7$ & $n=10$ & $n=15$\\ [6pt]\hline

& & \multicolumn{11}{c}{$\rho_0=-0.4,\rho_1=-0.2$} \\[4pt]\hline
$0.90$ &  & $ 24.2$ & $ 13.1$ & $  9.8$ & $  6.6$ & $  4.0$ &  & $ 29.1$ & $ 21.6$ & $ 17.6$ & $ 12.3$ & $  6.8$ \\
$0.95$ &  & $ 28.3$ & $ 25.2$ & $ 23.8$ & $ 21.7$ & $ 18.5$ &  & $ 29.6$ & $ 27.8$ & $ 26.8$ & $ 25.3$ & $ 22.6$ \\
$0.98$ &  & $ 29.6$ & $ 29.0$ & $ 28.7$ & $ 28.4$ & $ 27.8$ &  & $ 29.9$ & $ 29.4$ & $ 29.3$ & $ 29.0$ & $ 28.5$ \\
$0.99$ &  & $ 29.9$ & $ 29.6$ & $ 29.5$ & $ 29.4$ & $ 29.3$ &  & $ 30.0$ & $ 29.8$ & $ 29.7$ & $ 29.6$ & $ 29.5$ \\
$1.00$ &  & $ 30.0$ & $ 30.1$ & $ 30.1$ & $ 30.1$ & $ 30.1$ &  & $ 30.1$ & $ 30.1$ & $ 30.1$ & $ 30.1$ & $ 30.1$ \\
\hline
\end{tabular}
}
\label{tab:Table12}
\end{table}
\end{flushleft}

\begin{flushleft}
\begin{table}[htbp]
\centering
\caption{$TARL_{1}$ values of the one-sided Shewhart $RZ$ chart ($Sh^{-}_{RZ}$ and $Sh^{+}_{RZ}$) for $z_0=1, \gamma_{X}\in \left\{0.01, 0.2\right\}, \gamma_{Y} \in \left\{0.01, 0.2\right\}, \gamma_{X}\neq\gamma_{Y}, \rho_0 \in \left\{-0.8, -0.4, 0.0, 0.4, 0.8\right\}, \rho_1\neq\rho_0, n=\left\{1, 5, 7, 10, 15\right\}, \tau=\left\{0.9, 0.95, 0.98, 0.99, 1, 1.01, 1.02, 1.05, 1.1 \right\}$ and $TARL_{0}=30.$}
\resizebox{1.0\textwidth}{!}{
\begin{tabular}{l@{\hskip 0.2in}r@{\hskip 0.2in}r@{\hskip .2in}r@{\hskip .2in}r@{\hskip .2in}r@{\hskip .2in}r@{\hskip .2in}r@{\hskip .2in}r@{\hskip .2in}r@{\hskip .2in}r@{\hskip .2in}r@{\hskip .2in}r}\hline
& &\multicolumn{5}{c}{$(\gamma_X, \gamma_Y)=(0.01, 0.2)$} & \phantom{abc}& \multicolumn{5}{c}{$(\gamma_X, \gamma_Y)=(0.2, 0.01)$} \\
\cmidrule{3-7} \cmidrule{9-13} 
$\tau$ && $n=1$ & $n=5$ & $n=7$ & $n=10$ & $n=15$ &&$n=1$ & $n=5$ & $n=7$ & $n=10$ & $n=15$\\ [6pt]\hline

& & \multicolumn{11}{c}{$\rho_0=-0.4,\rho_1=-0.2$} \\[4pt]\hline
$0.90$ &  & $ 24.2$ & $ 13.1$ & $  9.8$ & $  6.6$ & $  4.0$ &  & $ 29.1$ & $ 21.6$ & $ 17.6$ & $ 12.3$ & $  6.8$ \\
$0.95$ &  & $ 28.3$ & $ 25.2$ & $ 23.8$ & $ 21.7$ & $ 18.5$ &  & $ 29.6$ & $ 27.8$ & $ 26.8$ & $ 25.3$ & $ 22.6$ \\
$0.98$ &  & $ 29.6$ & $ 29.0$ & $ 28.7$ & $ 28.4$ & $ 27.8$ &  & $ 29.9$ & $ 29.4$ & $ 29.3$ & $ 29.0$ & $ 28.5$ \\
$0.99$ &  & $ 29.9$ & $ 29.6$ & $ 29.5$ & $ 29.4$ & $ 29.3$ &  & $ 30.0$ & $ 29.8$ & $ 29.7$ & $ 29.6$ & $ 29.5$ \\
$1.00$ &  & $ 30.0$ & $ 30.1$ & $ 30.1$ & $ 30.1$ & $ 30.1$ &  & $ 30.1$ & $ 30.1$ & $ 30.1$ & $ 30.1$ & $ 30.1$ \\
\hline
\end{tabular}
}
\label{tab:Table13}
\end{table}
\end{flushleft}
\begin{flushleft}
\begin{table}[htbp]
\centering
\caption{$TARL_{1}$ values of the one-sided Shewhart $RZ$ chart ($Sh^{-}_{RZ}$ and $Sh^{+}_{RZ}$) for $z_0=1, \gamma_{X}\in \left\{0.01, 0.2\right\}, \gamma_{Y} \in \left\{0.01, 0.2\right\}, \gamma_{X}=\gamma_{Y}, \rho_0 \in \left\{-0.8, -0.4, 0.0, 0.4, 0.8\right\}, \rho_1\neq\rho_0, n=\left\{1, 5, 7, 10, 15\right\}, \tau=\left\{0.9, 0.95, 0.98, 0.99, 1, 1.01, 1.02, 1.05, 1.1 \right\}$ and $TARL_{0}=50.$}
\resizebox{1.0\textwidth}{!}{
\begin{tabular}{l@{\hskip 0.2in}r@{\hskip 0.2in}r@{\hskip .2in}r@{\hskip .2in}r@{\hskip .2in}r@{\hskip .2in}r@{\hskip .2in}r@{\hskip .2in}r@{\hskip .2in}r@{\hskip .2in}r@{\hskip .2in}r@{\hskip .2in}r}\hline
& &\multicolumn{5}{c}{$(\gamma_X, \gamma_Y)=(0.01, 0.01)$} & \phantom{abc}& \multicolumn{5}{c}{$(\gamma_X, \gamma_Y)=(0.2, 0.2)$} \\
\cmidrule{3-7} \cmidrule{9-13} 
$\tau$ && $n=1$ & $n=5$ & $n=7$ & $n=10$ & $n=15$ &&$n=1$ & $n=5$ & $n=7$ & $n=10$ & $n=15$\\ [6pt]\hline

& & \multicolumn{11}{c}{$\rho_0=-0.4,\rho_1=-0.2$} \\[4pt]\hline
$0.90$ &  & $ 41.9$ & $ 22.8$ & $ 16.5$ & $ 10.6$ & $  6.0$ &  & $ 49.1$ & $ 39.4$ & $ 32.8$ & $ 23.3$ & $ 12.3$ \\
$0.95$ &  & $ 47.9$ & $ 43.6$ & $ 41.5$ & $ 38.3$ & $ 33.0$ &  & $ 49.7$ & $ 47.5$ & $ 46.3$ & $ 44.3$ & $ 40.4$ \\
$0.98$ &  & $ 49.6$ & $ 48.8$ & $ 48.5$ & $ 48.1$ & $ 47.4$ &  & $ 49.9$ & $ 49.4$ & $ 49.2$ & $ 48.9$ & $ 48.4$ \\
$0.99$ &  & $ 49.9$ & $ 49.6$ & $ 49.5$ & $ 49.4$ & $ 49.2$ &  & $ 50.0$ & $ 49.8$ & $ 49.7$ & $ 49.6$ & $ 49.4$ \\
$1.00$ &  & $ 50.0$ & $ 50.1$ & $ 50.1$ & $ 50.1$ & $ 50.1$ &  & $ 50.2$ & $ 50.1$ & $ 50.1$ & $ 50.1$ & $ 50.1$ \\
\hline
\end{tabular}
}
\label{tab:Table14}
\end{table}
\end{flushleft}

\begin{flushleft}
\begin{table}[htbp]
\centering
\caption{$TARL_{1}$ values of the one-sided Shewhart $RZ$ chart ($Sh^{-}_{RZ}$ and $Sh^{+}_{RZ}$) for $z_0=1, \gamma_{X}\in \left\{0.01, 0.2\right\}, \gamma_{Y} \in \left\{0.01, 0.2\right\}, \gamma_{X}\neq\gamma_{Y}, \rho_0 \in \left\{-0.8, -0.4, 0.0, 0.4, 0.8\right\}, \rho_1\neq\rho_0, n=\left\{1, 5, 7, 10, 15\right\}, \tau=\left\{0.9, 0.95, 0.98, 0.99, 1, 1.01, 1.02, 1.05, 1.1 \right\}$ and $TARL_{0}=50.$}
\resizebox{1.0\textwidth}{!}{
\begin{tabular}{l@{\hskip 0.2in}r@{\hskip 0.2in}r@{\hskip .2in}r@{\hskip .2in}r@{\hskip .2in}r@{\hskip .2in}r@{\hskip .2in}r@{\hskip .2in}r@{\hskip .2in}r@{\hskip .2in}r@{\hskip .2in}r@{\hskip .2in}r}\hline
& &\multicolumn{5}{c}{$(\gamma_X, \gamma_Y)=(0.01, 0.2)$} & \phantom{abc}& \multicolumn{5}{c}{$(\gamma_X, \gamma_Y)=(0.2, 0.01)$} \\
\cmidrule{3-7} \cmidrule{9-13} 
$\tau$ && $n=1$ & $n=5$ & $n=7$ & $n=10$ & $n=15$ &&$n=1$ & $n=5$ & $n=7$ & $n=10$ & $n=15$\\ [6pt]\hline

& & \multicolumn{11}{c}{$\rho_0=-0.4,\rho_1=-0.2$} \\[4pt]\hline
$0.90$ &  & $ 41.9$ & $ 22.8$ & $ 16.5$ & $ 10.6$ & $  6.0$ &  & $ 49.1$ & $ 39.4$ & $ 32.8$ & $ 23.3$ & $ 12.3$ \\
$0.95$ &  & $ 47.9$ & $ 43.6$ & $ 41.5$ & $ 38.3$ & $ 33.0$ &  & $ 49.7$ & $ 47.5$ & $ 46.3$ & $ 44.3$ & $ 40.4$ \\
$0.98$ &  & $ 49.6$ & $ 48.8$ & $ 48.5$ & $ 48.1$ & $ 47.4$ &  & $ 49.9$ & $ 49.4$ & $ 49.2$ & $ 48.9$ & $ 48.4$ \\
$0.99$ &  & $ 49.9$ & $ 49.6$ & $ 49.5$ & $ 49.4$ & $ 49.2$ &  & $ 50.0$ & $ 49.8$ & $ 49.7$ & $ 49.6$ & $ 49.4$ \\
$1.00$ &  & $ 50.0$ & $ 50.1$ & $ 50.1$ & $ 50.1$ & $ 50.1$ &  & $ 50.2$ & $ 50.1$ & $ 50.1$ & $ 50.1$ & $ 50.1$ \\
\hline
\end{tabular}
}
\label{tab:Table15}
\end{table}
\end{flushleft}
Tables \ref{tab:Table10}-\ref{tab:Table15} indicate   the $TARL_{1}$ of two  ${\mathcal{S}h}^-_{RZ}$ and ${\mathcal{S}h}^+_{RZ}$ charts for $I=10,\ 30$ and $50$, when $\rho_1\neq \rho_0$ using $ \ n\in \left\{1,5,7,15,30\right\},\ \gamma_{X}\in \left\{0.01,0.2\right\}, \ \gamma_{Y}\in \left\{0.01,\ 0.2\right\}$ and $\tau=\{0.9,0.95,.98, 0.99,1.01,1.02,1.05,1.1\}$. Result discussion can be summarized of Tables \ref{tab:Table10}-\ref{tab:Table15} as follows
\begin{enumerate}
\item  Charts showed the asymmetrical performance due to biased feature of $TARL_{1}$ for all values of $n,   I,  \gamma_{X}$ and $\gamma_{Y}$. Since in control  $I=10, 30$, and $50$ are not stable. 

\item  It is the worth of observing that when $\gamma_{X}<\gamma_{Y}$ and $\rho_1 = \rho_0$, the one sided ${\mathcal{S}h}^-_{RZ}$ chart showed the sensitivity towards shifts. Which is conversely true for one sided ${\mathcal{S}h}^+_{RZ}$ chart when $\gamma_{Y}$$<$$\gamma_{X}$. For example, when $n=1,\ \rho_0=-0.4,\ \rho_1=\frac{\rho_0}{2}=-0.2$,  $I=30,\ \gamma_{X}=0.01,\ \gamma_{Y}=0.2,\ \tau=.99\ \&\ 1.05$, the values of TARL are $9.9$ and $10$. However, for $\gamma_{X}=0.2$\ and\ $\gamma_{Y}=0.01$ TARL are $10$ and $9.9$ (cf.  Table (\ref{tab:Table10}-\ref{tab:Table15}).
\end{enumerate}
\section{Real Life Illustration}
\label{sec:Real Life Illustration}
The following example has been introduced in \citet{Celano2014a} for the
implementation of a two-sided Shewhart control chart for monitoring the ratio of two normal variables in a long production run context
and will be adapted, in this paper, to a short production run context. This example considers a real quality
control problem from the food industry but with the simulation data. A muesli brand recipe is
produced by using a mixture of several ingredients including sunflower
oil, wildflower honey, seeds (pumpkin, flaxseeds, sesame, poppy),
coconut milk powder and rolled oats. To meet the food's nutrient
composition requirements declared in the brand packaging label and to
preserve the mixture taste, the recipe calls for equal weights of
``pumpkin seeds'' and ``flaxseeds''. Furthermore, their nominal
proportions to the total weight of box content are both fixed at
$p_p=p_f=0.1$. To satisfy the needs of customers, the brand boxes are
manufactured in different sizes. According to
\citet{Celano2014a}, whichever is the package dimension, the
quality practitioner wants to perform on-line SPC monitoring at
regular intervals $i=1,2,\ldots$ to check deviations from the
in-control ratio $z_0=\frac{\mu_{p,i}}{\mu_{f,i}}=\frac{0.1}{0.1}=1$
due to problems occurring at the dosing machine. Here, $\mu_{p,i}$ and
$\mu_{f,i}$ are the mean weights for ``pumpkin seeds'' and
``flaxseeds'', respectively.  The quality practitioner collects a
sample of $n=5$ boxes every 60 minutes. Because the box size is
allowed to change from one sample to another, it is possible to have
$\mu_{p,i}\neq\mu_{p,k}$ and $\mu_{f,i}\neq\mu_{f,k}$, $\forall i\neq
k$.  In the quality control laboratory, a mechanical procedure
separates the ``pumpkin seeds'' and ``flaxseeds'' from the muesli
mixture filling each box and the sample average weights
$\bar{W}_{p,i}=\frac{1}{n}\sum_{j=1}^nW_{p,i,j}$ and
$\bar{W}_{f,i}=\frac{1}{n}\sum_{j=1}^nW_{f,i,j}$ are
recorded. Finally, the ratio
$\hat{Z}_i=\frac{\bar{W}_{p,i}}{\bar{W}_{f,i}}$ is computed. 
For this example, like in
\citet{Celano2014a}, for $i=1,2,\ldots,$ and $j=1,2,\ldots,n$
both $W_{p,i,j}$ and $W_{f,i,j}$ can be well approximated as normal
variables with constant coefficients of variation $\gamma_p=0.02$ and
$\gamma_f=0.01$, i.e. $W_{p,i,j}\sim N(\mu_{p,i},0.02\times\mu
_{p,i})$ and $W_{f,i,j}\sim
N(\mu_{f,i},0.01\times\mu_{f,i})$. Moreover, the in-control
correlation coefficient between these two variables is
$\rho_0=0.8$. The process engineer has decided to implement   the ${\mathcal{S}h}^+_{RZ}$ chart   in order to monitor the ratio for a short run production of  $H=16$ hours calling for
$I=15$ inspections, i.e. an inspection every hour.
\begin{flushleft}
\begin{table}[htbp]
  \begin{center}
	\caption{The food industry example data}
\scalebox{0.68}{
  \begin{tabular}{ccccccccc}
    \hline
    & & \multicolumn{5}{c}{$W_{p,i,j}$ [gr]} & $\bar{W}_{p,i}$ [gr] &\\
    Sample & Box Size & \multicolumn{5}{c}{$W_{f,i,j}$ [gr]} & $\bar{W}_{f,i}$ [gr] & $\hat{Z_i}=\frac{\bar{W}_{p,i}}{\bar{W}_{f,i}}$\\
    \hline
    1  & 250 gr  &  $25.479$ & $25.355$ & $ 24.027$ & $25.792$ & $24.960$ &  $25.122$ & $1.003$ \\
           &         &  $25.218$ & $25.171$ & $ 24.684$ & $25.052$ & $25.107$ &  $25.046$ &  \\
    2  & 250 gr  &  $25.359$ & $25.172$ & $ 24.508$ & $25.292$ & $24.449$ &  $24.956$ & $1.003$ \\
    &                &  $25.211$ & $25.115$ & $ 24.679$ & $24.933$ & $24.831$ &  $24.954$ & \\[2mm]
    3  & 250 gr  &  $24.574$ & $24.864$ & $ 25.865$ & $25.107$ & $24.811$ &  $25.044$ & $1.005$ \\
    &                &  $24.784$ & $24.868$ & $ 25.377$ & $24.879$ & $24.734$ &  $24.929$ & \\[2mm]
    4  & 250 gr  &  $25.313$ & $24.483$ & $ 24.088$ & $25.184$ & $25.681$ &  $24.950$ & $0.999$ \\
    &                &  $25.338$ & $24.859$ & $ 24.305$ & $25.115$ & $25.251$ &  $24.974$ & \\[2mm]
    5  & 250 gr  &  $25.557$ & $24.959$ & $ 25.023$ & $24.482$ & $25.531$ &  $25.111$ & $0.998$ \\
    &                &  $25.277$ & $25.402$ & $ 25.012$ & $24.937$ & $25.148$ &  $25.163$ & \\[2mm]
    6  & 250 gr  &  $24.882$ & $24.473$ & $ 24.814$ & $25.418$ & $24.732$ &  $24.864$ & $0.997$\\
    &                &  $24.962$ & $24.644$ & $ 24.817$ & $25.419$ & $24.818$ &  $24.932$ & \\[2mm]
    7  & 500 gr  &  $49.848$ & $48.685$ & $ 49.994$ & $49.910$ & $49.374$ &  $49.562$ & $0.999$\\
    &                &  $49.993$ & $49.128$ & $ 49.830$ & $49.566$ & $49.422$ &  $49.588$ & \\[2mm]
    8  & 500 gr  &  $49.668$ & $50.338$ & $ 49.149$ & $47.807$ & $49.064$ &  $49.205$ & $0.990$\\
    &                &  $49.695$ & $50.681$ & $ 49.640$ & $48.969$ & $49.612$ &  $49.720$ & \\[2mm]  
    9  & 500 gr  &  $51.273$ & $48.303$ & $ 48.510$ & $50.594$ & $48.591$ &  $49.454$ & $0.993$ \\
    &                &  $50.366$ & $49.210$ & $ 49.844$ & $49.890$ & $49.595$ &  $49.781$ & \\[2mm]
    10  & 500 gr  &  $48.720$ & $51.566$ & $ 49.677$ & $50.651$ & $50.344$ &  $50.192$ & $1.002$\\
    &                &    $49.721$ & $50.215$ & $ 50.178$ & $50.324$ & $50.071$ &  $50.102$ & \\[2mm]
    11  & 500 gr  &  $51.372$ & $51.700$ & $ 51.000$ & $50.886$ & $49.641$ &  $50.920$ & ${\bf 1.017}$\\
    &                &    $50.164$ & $50.272$ & $ 49.884$ & $50.061$ & $49.845$ &  $50.045$ & \\[2mm]
    12  & 500 gr  &  $52.020$ & $53.182$ & $ 51.374$ & $51.342$ & $48.771$ &  $51.138$ & ${\bf 1.023}$\\
    &                &    $50.749$ & $50.369$ & $ 49.697$ & $49.575$ & $49.440$ &  $49.966$ & \\[2mm]
    13  & 500 gr  &  $52.360$ & $49.412$ & $ 50.704$ & $50.370$ & $50.901$ &  $50.949$ & ${\bf 1.016}$\\
    &                &    $50.047$ & $49.981$ & $ 50.297$ & $50.408$ & $50.026$ &  $50.152$ & \\[2mm]
    14  & 500 gr  &  $52.498$ & $50.447$ & $ 48.713$ & $48.574$ & $50.275$ &  $50.101$ & $1.008$\\
    &                &    $50.064$ & $50.124$ & $ 49.162$ & $48.865$ & $50.344$ &  $49.712$ & \\[2mm]
    15  & 250 gr  &  $25.123$ & $24.658$ & $ 24.468$ & $25.030$ & $25.071$ &  $24.870$ & $0.996$\\
    &                &    $25.041$ & $24.790$ & $ 24.835$ & $25.211$ & $25.008$ &  $24.977$ & \\
    \hline
  \end{tabular}
  }
	\label{tab:Table16}
  \end{center}
\end{table}
\end{flushleft}

\begin{figure}
\begin{center}
    \includegraphics{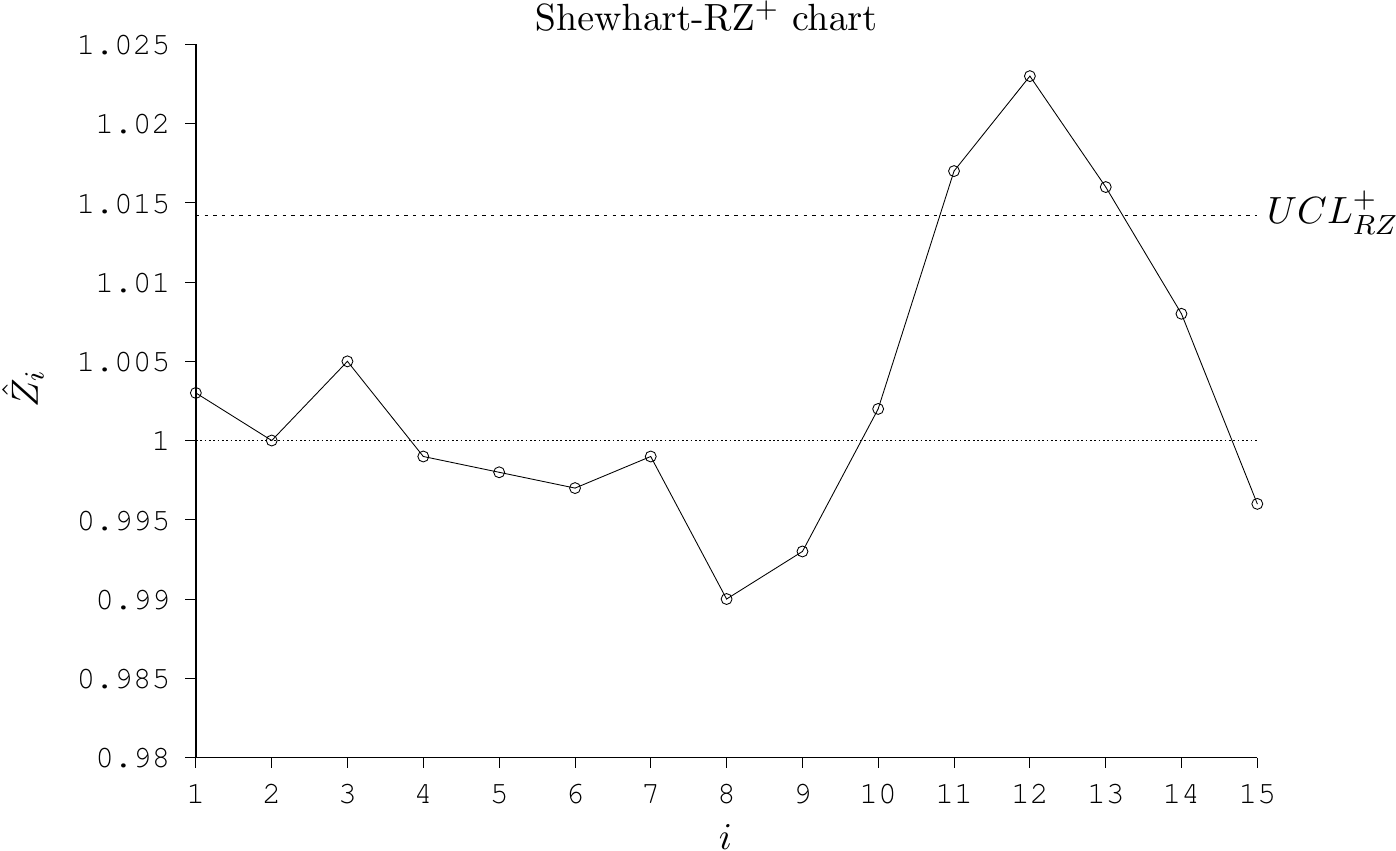} 
  \caption{  The ${\mathcal{S}h}^+_{RZ}$ control chart   for the food
    industry example}
\end{center}
\label{fig:Figure2}
\end{figure}

%

For $n=5$  and  $\rho_0=0.8$, the control limit of the  ${\mathcal{S}h}^+_{RZ}$ control chart  to potentially detect
unexpected increase in the ratio for a short run production is $UCL = 1.01421$. 
Table -\ref{tab:Table16} shows the
set of simulated sample data collected from the process (based on
\citet{Celano2014a}), the corresponding box sizes 250--500 gr,  and 
the $\hat{Z}_i$ statistic. The process is assumed to run
in-control up to sample \#10. Then, between samples \#10 and \#11,
\citet{Celano2014a} have simulated the occurrence of an
assignable cause shifting $z_0=1$ to $z_1=1.01\times z_0$, i.e. a
ratio percentage increase equal to 1\%. Figure 2 shows the
 ${\mathcal{S}h}^+_{RZ}$ control chart, which signals the occurrence of the
out-of-control condition by plotting point \#11 above the control
limit $UCL^+=1.01421$ (see also bold values in Table -\ref{tab:Table16}). 
The process is allowed to continue, while
corrective actions are started by the repair crew who find and
eliminate the assignable cause after sample \#14 and restore the
process back to the in-control condition.

\section{Conclusion and Recommendations}
\label{sec:Conclusion}
In this article we examined the statistical performance of two one-sided Shewhart charts $({\mathcal{S}h}^-_{RZ}$ and ${\mathcal{S}h}^+_{RZ}$) in monitoring the ratio of two normal variables using truncated average run length as a performance measure in short production runs. For practical consideration of quality experts, we have presented the probability limits and TARL values of both charts by varying the number of inspections. The sample size(n) and the number of inspections(M) affect the width of the control limits. Quality practitioners maybe adjust control limits according to the requirement of the production process. We observed the TARL performance of the both designs at fixed values $n, \rho_0,\ \rho_1$,  $\gamma_{X,\ }{\ \gamma}_{Y}, \tau\ $, $Z_0$ at pre-specified  ${TARL}_0=10,\ 30\ \&\ \ 50$ . The main conclusions maybe drive from results. The performance of chart one-sided Shewhart charts $({\mathcal{S}h}^-_{RZ}$ and ${\mathcal{S}h}^+_{RZ}$) is depend on the values $n,\ {\gamma}_{X},\ \gamma_{Y}$ and $\rho_0$. When shift size increase or decrease the values of TRL increase or decrease accordingly. There are several extensions for future research like one-sided  EWMA type and CUSUM type ratio chart for short production runs.

\bibliographystyle{unsrtnat}
\bibliography{Reference}

\end{document}